\documentclass[usenatbib]{mnras}

\usepackage[T1]{fontenc}
\usepackage{ae,aecompl}

\usepackage{graphicx}	
\usepackage{amsmath}	
\usepackage{amssymb}	

\title[Innate origin of gradients in a simulated galaxy disc]{The innate origin of
  radial and vertical gradients in a simulated
  galaxy disc}

\author[Navarro et al.]{Julio F. Navarro$^{1}$\thanks{E-mail:jfn@uvic.ca}, Cameron Yozin$^1$, Nic Loewen$^{1}$, 
  Alejandro Ben\'{i}tez-Llambay$^{2}$, \newauthor
  Azadeh Fattahi$^{1}$, 
 Carlos S. Frenk$^{2}$,
  Kyle A. Oman$^{1}$, 
Joop Schaye$^3$, 
Tom Theuns$^2$.
\\
$^{1}$Department of Physics and Astronomy, University of Victoria, Victoria, BC, Canada V8P 5C2\\
$^{2}$Institute for Computational Cosmology, Department of Physics, Durham University, South Road, Durham, DH1 3LE, UK\\
$^{3}$Leiden Observatory, Leiden University, PO Box 9513, 2300 RA Leiden, The Netherlands\\
}

\date{Accepted XXX. Received YYY; in original form ZZZ}

\pubyear{2016}

\begin{document}
\label{firstpage}
\pagerange{\pageref{firstpage}--\pageref{lastpage}}
\maketitle

\begin{abstract} 

  We examine the origin of radial and vertical gradients in the
  age/metallicity of the stellar component of a galaxy disc formed in
  the APOSTLE cosmological hydrodynamical simulations. Some of these
  gradients resemble those in the Milky Way, where they have sometimes
  been interpreted as due to internal evolution, such as scattering
  off giant molecular clouds, radial migration driven by spiral
  patterns, or orbital resonances with a bar. 
Secular processes
    play a minor role in the simulated galaxy, which lacks strong
    spiral or bar patterns, and where such gradients arise as a result
    of the gradual enrichment of a gaseous disc that is born thick but
    thins as it turns into stars and settles into centrifugal
    equilibrium.
 The settling is controlled by the feedback of young
  stars; which links the star formation, enrichment, and equilibration
  timescales, inducing radial and vertical gradients in the gaseous
  disc and its descendent stars. The kinematics of coeval stars evolve
  little after birth and provide a faithful snapshot of the gaseous
  disc structure at the time of their formation. In this
  interpretation, the age-velocity dispersion relation would reflect
  the gradual thinning of the disc rather than the importance of
  secular orbit scattering; the outward flaring of stars would result
  from the gas disc flare rather than from radial migration; and
  vertical gradients would arise because the gas disc gradually
  thinned as it enriched. Such radial and vertical trends might just
  reflect the evolving properties of the parent gaseous disc, and are
  not necessarily the result of secular evolutionary processes.
\end{abstract}





\section{Introduction}
\label{SecIntro}

The ages and metallicities of disc stars in the Milky Way (MW) show
well-established radial, vertical, and kinematic gradients. In the
solar neighbourhood, for example, the velocity dispersion of stars
increases with age \citep[e.g.,][]{Wielen1977,Quillen2001,Holmberg2009},
and age correlates in turn with total metallicity and, in particular, with
the abundance of $\alpha$ elements at given [Fe/H]
\citep[e.g.,][]{Freeman2002,Haywood2013}.  In the disc midplane, the average age and
metallicity of stars decline with increasing cylindrical radius, $R$
\citep[e.g.,][]{Boeche2013,Hayden2014}. The radial age trend is
preserved but the metallicity gradient appears to reverse when
considering stars at fixed vertical distances, $|z|$, away from the
disc symmetry plane \citep[e.g.,][]{Anders2014,Bergemann2014,Martig2016,Ness2016}.

At fixed $R$, the vertical scaleheight of stars increases with
decreasing metallicity or increasing age
\citep[e.g.,][]{Mikolaitis2014,Casagrande2016}; these gradients depend on $R$
and become steeper at smaller radii \citep{Hayden2014}.  Finally, at fixed metallicity,
the vertical scaleheight of stars increases (`flares') from the centre
outwards, especially for stars with low $[\alpha/$Fe$]$ ratios; the
flare depends on metallicity, becoming more pronounced at lower
[Fe/H] \citep{Bovy2016}.

Although the detailed form of each of these correlations vary somewhat
from study to study, their qualitative nature seems robust. Their
physical origin, however, is still a matter of debate. One possibility
is that, to first order, the differences between old (metal-poor)
stars and young (metal-rich) stars are driven by the integrated effect
of secular evolutionary processes, such as orbital scattering off
molecular clouds, radial migration induced by non-axisymmetric disc
features, or dynamical heating by accretion events, external
perturbations, and disc instabilities. These processes naturally
affect older stars more, inducing vertical gradients and systematic
trends between age/metallicity and kinematics that resemble the
observed ones \citep[for a review, see][and references
therein]{Rix2013}.

A contrasting view is that stellar gradients are largely imprinted at
birth, and thus reflect the properties of the parent (gaseous) disc at
the time of formation of each coeval stellar population. In this
scenario, the star formation and enrichment history of the disc are
linked to the evolution of its vertical structure and to the timescale
of its dynamical equilibration
\citep{Brook2012b,Bird2013,Stinson2013,Grand2015,Miranda2016,Ma2017,Grand2017}.
At early times the gaseous disc was thick and metal poor, but it
enriched as it thinned down, leaving behind a strong vertical gradient
in the age and metallicity of the stars at each fixed radius, in a
process reminiscent of the classic \citet{Eggen1962} dissipative
collapse picture. Radial gradients arise in this model simply because
the gaseous disc assembles inside out and is denser near the centre,
leading to faster transformation of gas into stars and more rapid
enrichment than in the outskirts.

Each of these views has its pros and cons. For example, secular
evolution models have difficulty accounting for the radially
increasing scaleheight (`flare') of stars, since kinematic
perturbations are expected to be weaker at larger radii on account of
lower densities, and to proceed more slowly there because of longer
orbital times. Birth models, on the other hand, have difficulty
explaining why the vertical settling and enrichment timescales needed
to explain the data are so much longer than the local dynamical
timescale, which would naively be expected to set the collapse and
equilibration timescale of the disc.

In the Milky Way, a further complication is the presence of an apparent `gap' in the
$\alpha$ vs Fe abundances of stars at fixed $R$, which has often been
interpreted as signalling the presence of two disc populations of
possibly distinct origin: a relatively old, $\alpha$-rich `thick'
disc mixed with a younger, $\alpha$-poor `thin' population of
overlapping [Fe/H] \citep[see; e.g.,][and references
therein]{Venn2004,Bensby2005,Navarro2011,Recio-Blanco2014}.

These complexities explain why `chemodynamical' models resort to
  features from both formation scenarios in order to reproduce the
  structure of the Milky Way disc(s).  Radial migration; external
  accretion; self-enrichment; gas outflows and inflows; these are some
  of the mechanisms invoked to reproduce the peculiar kinematic and
  chemical structure of the Galactic thin and thick discs \citep[see;
  e.g.,][and references
  therein]{Matteucci1989,Chiappini2001,Minchev2013,Minchev2014,Schoenrich2017}. The
  specificity of such models, however, hinders their general applicability to
  understanding the relative importance of various physical mechanisms
  in a typical disc galaxy.

Direct cosmological hydrodynamical simulations of disc galaxy
formation offer a complementary approach, where, modulo the
algorithmic choices for simulating star formation and feedback, the
roles of evolutionary effects and of conditions `at birth'
may be contrasted and assessed in statistically significant
samples. The price to pay is that very few, if any, of the models will
reproduce the detailed structure of the Milky Way, implying that the
physical origin of observed Galactic trends must be inferred from
judicious application of lessons learned from simulated galaxies that
might not necessarily resemble our Galaxy in detail.

We adopt this approach here, where we analyze the formation of a disc
galaxy in the APOSTLE\footnote{A Project Of Simulating The Local
  Environment} suite of cosmological hydrodynamical simulations of the
Local Group \citep{Sawala2016b,Fattahi2016}.  Our analysis focusses on the physical origin of vertical
and radial gradients of disc stars, on how they reflect the properties
of the gaseous disc from which they descend, and on the effect of
secular evolutionary processes that operate after their formation.
Our paper is organized as follows. We describe briefly the APOSTLE
numerical simulations in Sec.~\ref{SecNumSims}. Our main results are
presented in Sec.~\ref{SecResults}. A toy model meant to illustrate
how these results may be used to interpret the physical origin of
vertical gradients is presented in Sec.~\ref{SecModel}. We end with a
brief summary of our main conclusions in Sec.~\ref{SecConc}.

\section{Numerical Simulations}
\label{SecNumSims}

\subsection{The {\sc APOSTLE} project}
\label{SecAPOSTLE}

The {\sc APOSTLE} suite of zoomed-in $\Lambda$CDM cosmological
hydrodynamical simulations evolves 12 volumes from a large
cosmological box selected to resemble the main characteristics of the Local
Group (LG). Each volume is centred on a pair of dark matter halos of
combined virial\footnote{The virial properties of a dark matter halo
  are computed within spheres of mean density $200\times$ the critical
  density for closure, $\rho_{\rm crit}=3H_0^2/8\pi G$. Virial
  quantities will be indicated with a `200' subscript.} mass of order
$\approx 2\times 10^{12}\, M_\odot$. The pairs at lookback time $t=0$\footnote{We shall
  hereafter use lookback time, $t$, to quote simulation results, in
  order to prevent confusion between `redshift' and the vertical
  coordinate, $z$.} are at a relative distance of $\approx 800\pm 200$
kpc, approach with radial velocity of up to $\approx 250$ km/s, have
relative tangential velocities $<100$ km/s, and have no other halos of
comparable mass within $\approx 2.5$ Mpc. Details of the selection
procedure are presented in \citet{Sawala2016b,Fattahi2016}.

All {\sc APOSTLE} volumes were evolved with the same code developed
for the {\sc EAGLE} project \citep{Schaye2015, Crain2015}, which has
been calibrated to reproduce some general properties of the galaxy
population, such as the galaxy stellar mass function and
the average size of galaxies as a function of mass. {\sc APOSTLE}
extends the {\sc EAGLE} project to LG-like volumes and higher resolutions, using
the fiducial choice of parameters \citep[i.e., the `Ref' model in the nomenclature
of][]{Schaye2015}.

Each {\sc APOSTLE} volume is resimulated at three different numerical
resolutions, differing by successive factors of $\approx 10$ in particle
mass (labelled L1 to L3 in order of decreasing particle count). Our
analysis will focus on the most massive of the two main galaxies in volume AP-4-L2,
using the nomenclature introduced by \citet{Fattahi2016}. For that
run, the high-resolution particle masses are
$5.9\times 10^{5}\, M_\odot$ (dark matter) and
$1.2\times 10^{5}\, M_\odot$ (gas). These particles interact with a
gravitational potential softened with a Plummer-equivalent spline
kernel length fixed at $307$ pc for $z<3$ and constant in comoving
units at higher redshifts.

{\sc APOSTLE} runs assume a {\sc WMAP-7} cosmology throughout, with
parameters $\Omega_{\rm M}=0.272$, $\Omega_{\Lambda}=0.728$,
$\Omega_{\rm b}=0.0455$, $h=0.704$ and $\sigma_{8}=0.81$
\citep{Komatsu2011}. 

\begin{figure}
  \hspace{-0.2cm}
  \resizebox{8.5cm}{!}{\includegraphics{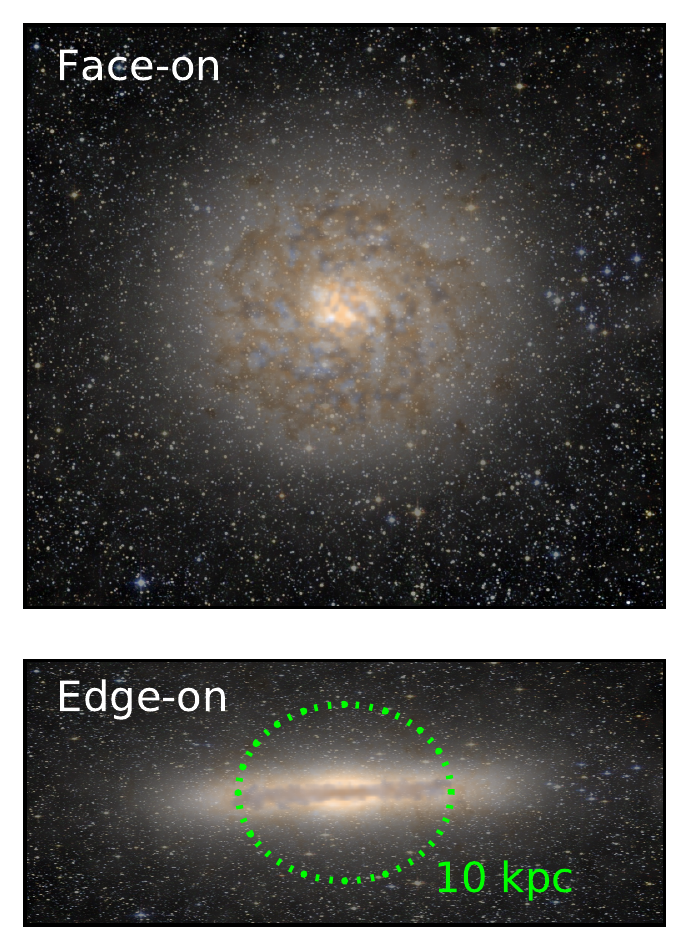}}\\%
  \caption{Mock image of the simulated galaxy at $t=0$ seen face-on and
    edge-on. False colours use $g,r,i$ broad bandpasses, as
  well as a simple prescription for dust absorption based on gas
  density and metallicity. The image is superposed on a artificial stellar field
  that mimics an image from the Hubble Space Telescope.}
\label{FigGxImage}
\end{figure}

\begin{figure}
  \hspace{-0.2cm}
  \resizebox{8.5cm}{!}{\includegraphics{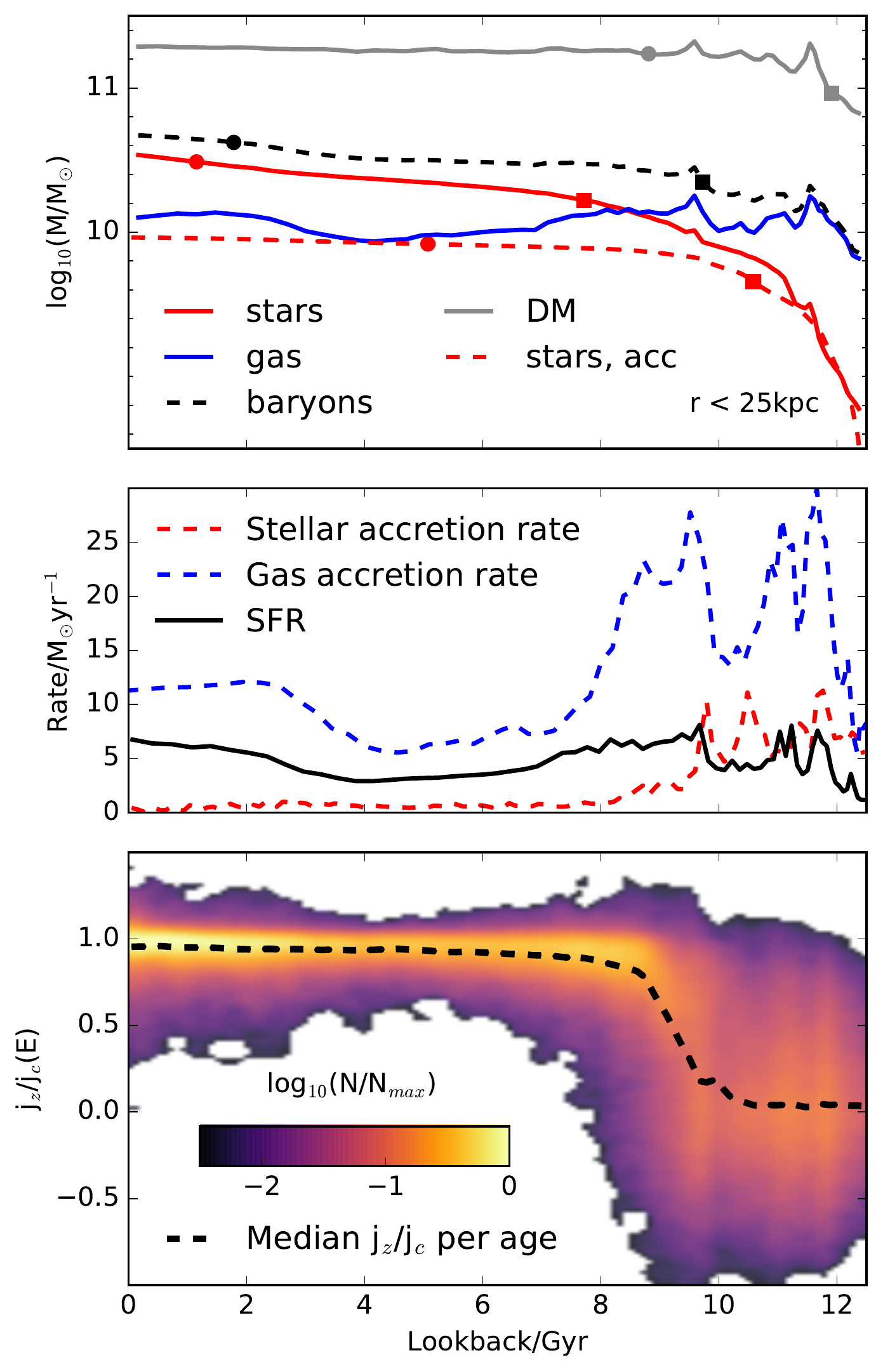}}\\%
  \caption{{\it Top panel:} Mass of various components of the galaxy
    within a fixed radius of $25$ (physical) kpc from the galaxy
    centre, as a function of lookback time. Mass components include
    dark matter (grey solid line), baryons (black dashed), stars (red
    solid), accreted stars (red dashed), and gas (blue solid). The squares
    and circles in each curve indicate the time when, respectively,
    $50\%$ and $90\%$ of the final mass was in place. {\it Middle
      panel:} Gas (blue) and stellar (red) accretion rates, as
    a function of lookback time. The solid black line indicates the
    in-situ star formation rate. {\it Bottom panel:} Stellar age
    vs. circularity ($\epsilon_J=j_z/j_{\rm circ}(E)$) for all stellar
    particles in the galaxy at $t=0$. Each pixel of the heat map is
    coloured according to a logarithmic function of particle
    number. Note that some particles have $\epsilon_J$ that slightly
    exceed unity; these are particles with $|z|>0$ whose energies are
    compared with those of the midplane circular orbit at the same
    $R$. The black dashed line indicates the median circularity.}
  \label{FigMassEvol}
\end{figure}

\begin{figure}
  \hspace{-0.2cm}
  \resizebox{8.5cm}{!}{\includegraphics{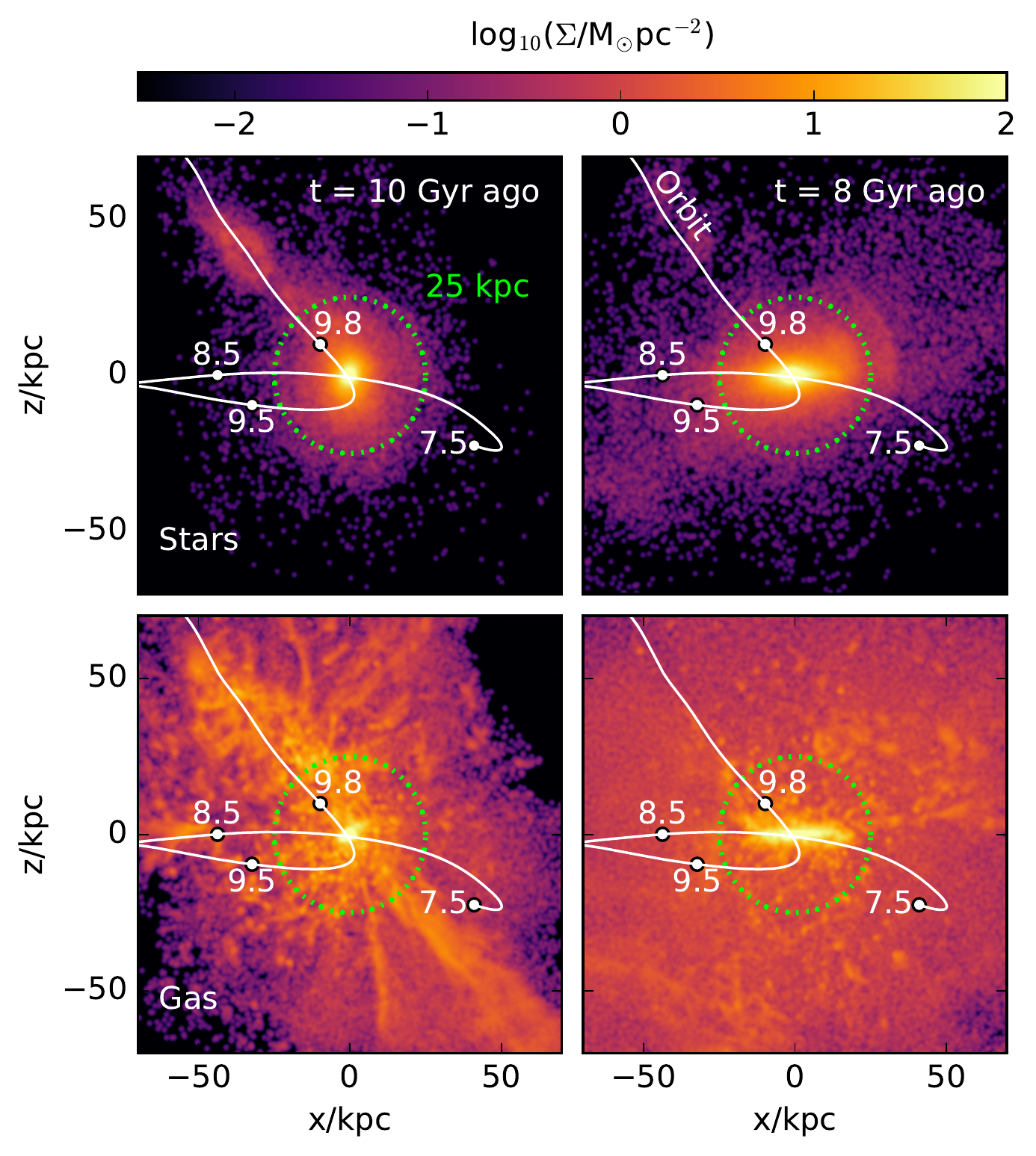}}\\%
  \caption{{\it Top row:} Projected distribution of stellar mass at
    two selected lookback times. The white solid line indicates the
    trajectory of the last large accreted satellite (a minor merger); labels
    indicate the location at given lookback time. {\it Bottom:}  As top row,
    but for the gaseous component. The accreted satellite is part of a
    filamentary structure that accreted roughly at the same time,
    substantially increasing the gas mass of the galaxy. The satellite
    trajectory is as in the top row.}
  \label{FigPertEvol}
\end{figure}

\subsection{The simulated galaxy}
\label{SecSimGx}

Dark matter halos are identified using a friends-of-friends
\citep[FoF;][]{Davis1985} algorithm run with linking length 0.2 times
the mean interparticle separation. Individual self-bound structures
within these halos are then identified recursively by the groupfinding
algorithm {\sc SUBFIND} \citep{Springel2001b,Dolag2009}. The galaxies
inhabiting the main subgroup of each FoF halo are referred to as
`centrals' and the rest as its `satellites', if found within the
virial radius of the central halo.

As stated above, we focus our analysis on a single galaxy, the more
massive of the pair of central galaxies in {\sc APOSTLE} volume
AP-4-L2.  This galaxy was selected because, at $t=0$, it (i) has a
prominent stellar disc; (ii) shows no obvious morphological
peculiarities; (iii) has formed stars steadily throughout its history;
(iv) has had a relatively quiet recent merging history; and (v) has a
gas/stellar mass ratio not unlike that of the Milky Way. We emphasize
that this galaxy is {\it not} a model of the Milky Way.  The
  properties of this galaxy are representative of other disc-dominated
  galaxies in APOSTLE selected using similar criteria; however,
  since our analysis involves the detailed tracking of individual star
  particles it is easier to illustrate our main results by focussing
  on a single galaxy.  Fig.~\ref{FigGxImage} shows face-on and
edge-on views of the galaxy, in a box $\approx 50$ kpc on a side.

\begin{figure*}
  \hspace{-0.2cm}
  \resizebox{17.8cm}{!}{\includegraphics{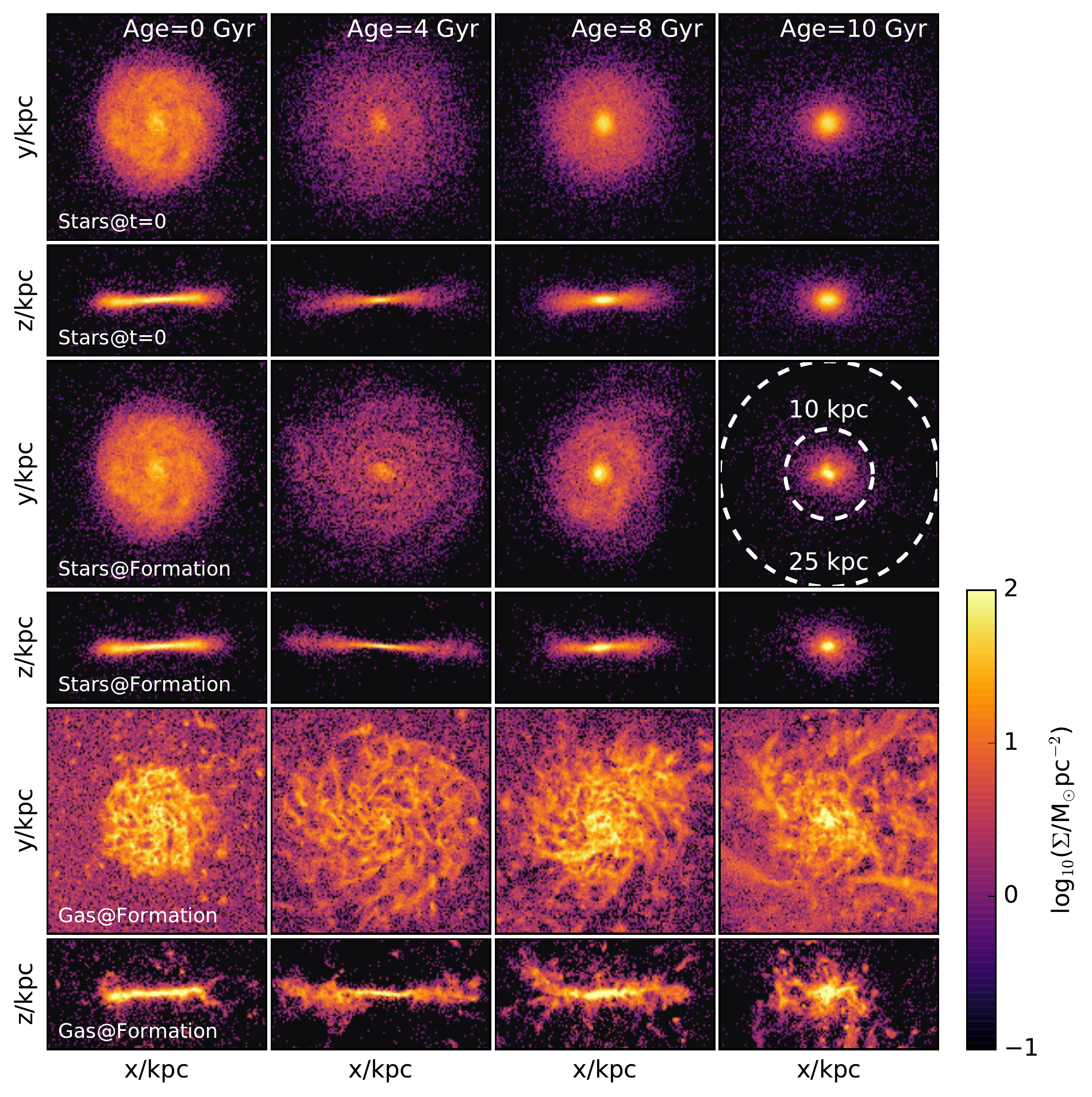}}\\%
  \caption{Face-on and edge-on projections of star and gas particles at
    selected times. Top two rows show, at $t=0$, stars with ages
    ($\pm 0.5$ Gyr) given in the legend of each box. Middle rows show
    the same star particles as in the top rows, but as they were at the time of
    formation. Bottom rows indicate the gas component at each of the
    selected times. Note how the gas slowly assembles into a gradually
    thinning disc largely broken up into star forming clumps.}
  \label{FigGxEvol}
\end{figure*}

\section{Results}
\label{SecResults}

\begin{figure}
  \hspace{-0.2cm}
  \resizebox{8.5cm}{!}{\includegraphics{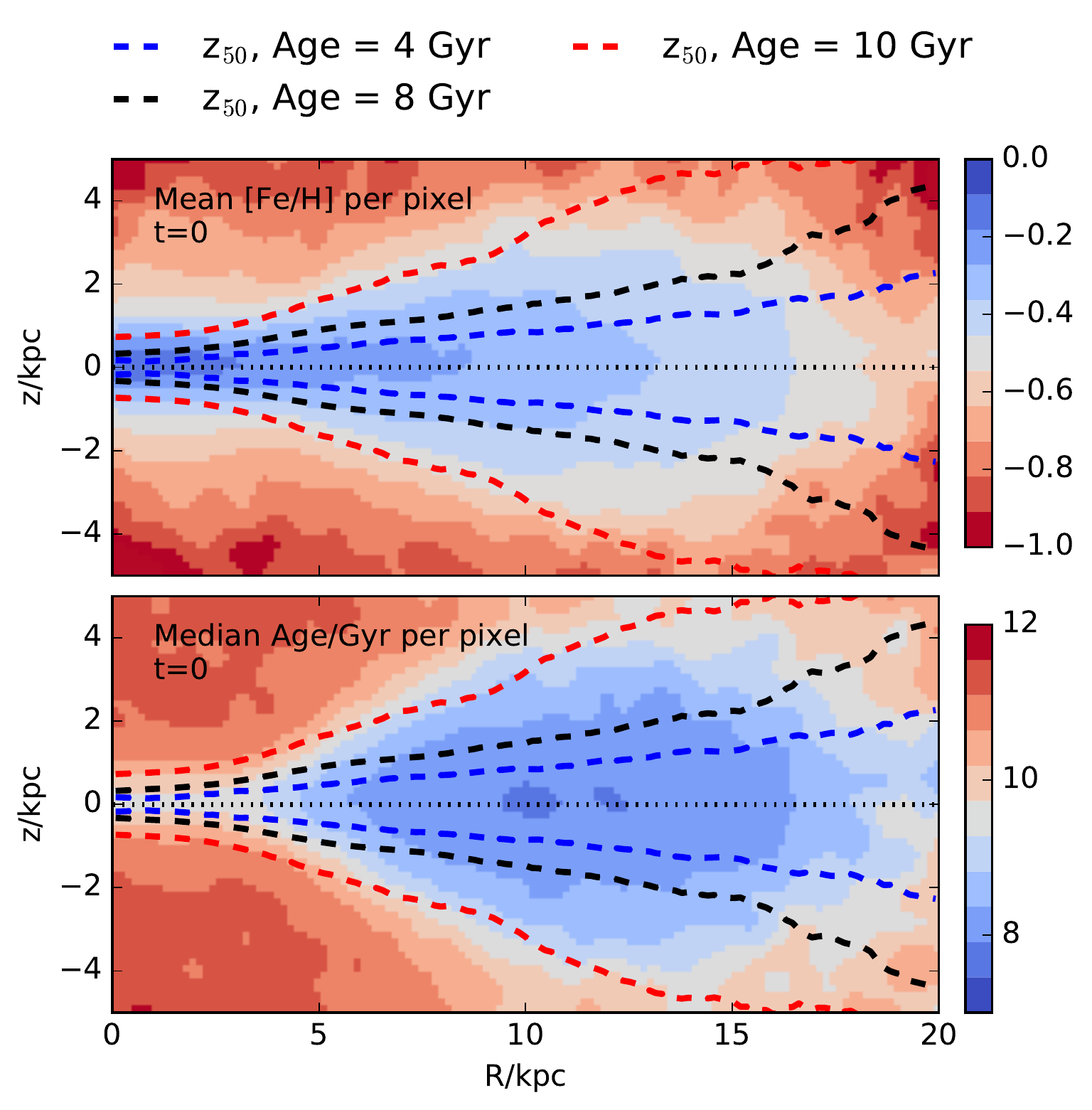}}\\%
  \caption{Average stellar metallicity (top) and age (bottom) of star particles
    at the present time, as a function of the cylindrical coordinates
    $R$ and $z$ (colour labels on right). Note that the disc is
    strongly flared, and that younger star particles delineate a thinner
    structure than their older counterparts. Dashed curves with
    different colours in both
    panels indicate the half-mass scaleheight of stars of ages $10$,
    $8$ and $4$ Gyr. The origin of the strong radial and vertical
    gradients seen in this figure is the main focus of this paper.}
  \label{FigAgeMetRGrad}
\end{figure}

\begin{figure}
  \hspace{-0.2cm}
  \resizebox{8.5cm}{!}{\includegraphics{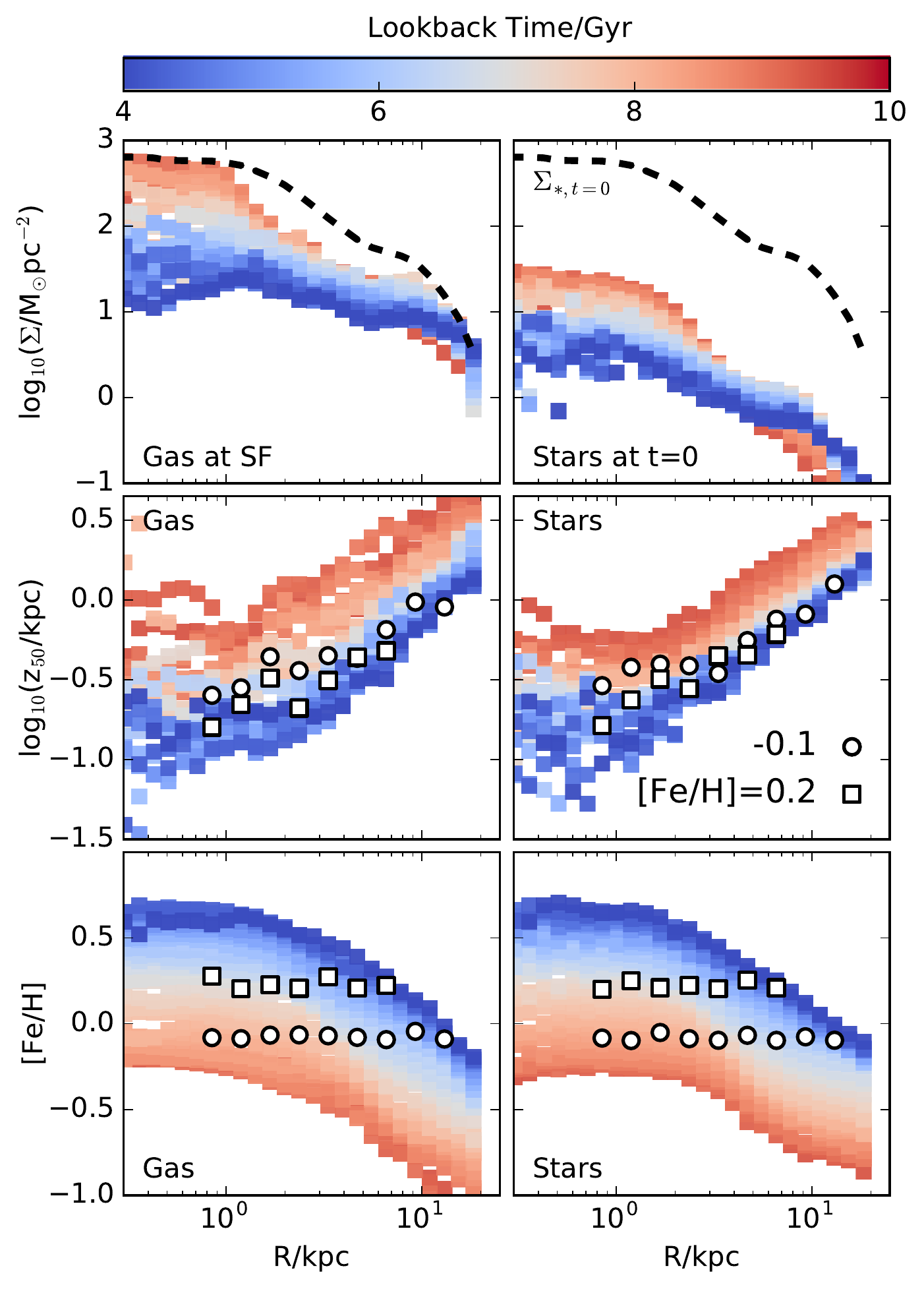}}\\%
  \caption{{\it Left column:}  Surface density (top), half-mass
    scaleheight (middle) and average metallicity (bottom) radial
    profiles of the gas component of the simulated galaxy, at various
    times, as indicated by the colour bar at the top. Note that the gas
    gradually enriches itself as it settles vertically and makes
    stars. The thick dashed line in the top panel indicates the
    surface density profile of all star particles at the present time,
    $t=0$. {\it Right column:} Same as left, but for star particles at $t=0$,
    grouped by age. Note the tight correspondance between
    stellar gradients as a function of age and those of the gas at the
    time of their formation. Open circles and squares highlight
    populations of fixed metallicity, [Fe/H]$=-0.1$ and $+0.2$,
    respectively. (See Fig.~\ref{FigMetZGrad} for further discussion.)}
  \label{FigGasStarGrad}
\end{figure}

\subsection{The assembly of the galaxy}
\label{SecGxAss}

The top panel of Fig.~\ref{FigMassEvol} shows the mass assembly
history of the galaxy, as measured by the total mass of each component
within a radius of $25$ kpc (physical) from the centre of the
progenitor halo. Most of the galaxy mass within that radius is
assembled quickly: $50\%$ of the dark matter mass was already in place
$11.5$ Gyr ago; the corresponding lookback time for the baryonic component is
$t\approx 9.5$ Gyr.

Most baryons are accreted as gas and then turn gradually into stars,
as shown in the middle panel of Fig.~\ref{FigMassEvol}. Gas is
delivered to the centre of the galaxy during an early period of rapid
merging that is essentially over $\approx 9$ Gyr ago. The last episode of
early accretion is depicted in Fig.~\ref{FigPertEvol}, where we see
that  the accretion of a substantial amount of gas along a filamentary
structure accompanies a minor merger that happened roughly $10$ Gyr
ago.

After that accretion event the evolution of the galaxy is largely
quiescent, except for an episode of gas accretion that started roughly
$3$ Gyr ago. This late accretion adds $\approx 10^{10}\, M_\odot$ of
gas, but not many stars. This addition increases the final baryonic
mass of the galaxy by $\approx 40\%$, but nearly doubles its gas mass
at the time of accretion. The extra gas (which is of relatively
  low metallicity) serves to ramp up the star formation rate of the
  galaxy and to decrease its average metallicity at late times.

The middle panel of Fig.~\ref{FigMassEvol} also makes clear that the
large majority of stars in this galaxy were born in situ\footnote{At
  very early times, when only a small fraction of the final galaxy has
  been assembled, the main progenitor is not well defined. This leads
  to temporary uncertainties in the assignment of in-situ vs accreted
  stars. These uncertainties have no discernible consequences on our
  main conclusions.}: fewer than $\approx 25\%$ of all star particles at $t=0$
formed in other systems, and roughly $90\%$ of those were accreted
more than $\approx 9$ Gyr ago.

The quiet late merging history of the galaxy, together with the fact
that most baryons are accreted in gaseous form, favour the formation of
a centrifugally supported disc, as may be seen in the bottom panel of
Fig.~\ref{FigMassEvol}. This panel shows the circularity parameter of
star particles, $\epsilon_J=j_z/j_{\rm circ}$, measured at $t=0$, as a function
of their formation time. The parameter $\epsilon_J$ is defined as the
$z$-component of the specific angular momentum of a star in units of
that of a circular orbit of the same binding energy, and it therefore
varies between $\pm 1$ for stars in the midplane. Some stars may have
$\epsilon_J$ slightly exceeding unity if they are outside the disc
midplane. The $z$ axis of the disc is defined at all times as the
direction of the angular momentum vector of star particles younger than $3$ Gyr
old.

Star particles older than $10$ Gyr formed before the early merging period of
the galaxy was over: indeed, $80\%$ of them are accreted and all
together they have no net sense of rotation. Over the next $\approx 2$
Gyr, however, a disc gradually forms and the average circularity
climbs to unity.  Stars formed during that epoch (i.e., ages between $8$
and $10$ Gyr) have a broad distribution of circularities, and the
great majority (more than $70\%$) formed in-situ. Stars younger than
$\approx 7$ Gyr, on the other hand, nearly all formed in-situ and are
found in approximately circular orbits in a thin coplanar disc (i.e.,
$\epsilon_J\approx 1$).

The various panels of Fig.~\ref{FigGxEvol} show a few
snapshots\footnote{This figure was inspired by Fig.~1 of
  \citet{Ma2017}.} of the evolution of the galaxy. Here, from right to
left, each column shows different orthogonal projections of star particles
formed, within a narrow interval of time, $10$, $8$, $4$, and $0$ Gyr
ago, respectively. The first two rows show each coeval population of
stars at $t=0$. The third and fourth rows, on the other
hand, show them at the time of their formation. The gas component of
the galaxy is shown in the bottom two rows, for each time.

Fig.~\ref{FigGxEvol} shows that, as expected, star particles trace, at the time
of formation, the densest regions of the gaseous disc. Interestingly,
however, the spatial configuration of stars evolves little after
formation, and remains similar at $t=0$. The stars of the simulated
disc, therefore, seem to provide {\it today} a relatively faithful tracer of the
properties of the gaseous disc at the time of their formation. This is
the central result of our study, and will be a recurrent theme in the
analysis that follows.

\subsection{Age and metallicity gradients}
\label{SecGrads}

The evolution of the gaseous disc discussed in the previous subsection
leaves discernible gradients in the age and metallicity of its
descendent stars at $t=0$. We show this in Fig.~\ref{FigAgeMetRGrad},
where the heat maps show, as a function of the cylindrical coordinates,
$R$ and $z$, the average age and metallicity of star particles. The gradients
shown are characteristic of the `inside-out/upside-down'
disc formation scenario described in earlier work \citep[see,
e.g.,][]{Bournaud2009,Brook2012b,Bird2013,Stinson2013,Miranda2016,Ma2017}.

In the disc midplane the average age and metallicity decrease
monotonically with increasing $R$. At fixed $R$, the
ages of star particles increase and their metallicities decrease with increasing
$|z|$. These gradients are steeper near the centre of the galaxy than
in its outskirts, implying that the characteristic scaleheight of
stars of fixed metallicity (or age) `flares' outwards. This
flare is the reason why the radial metallicity gradient at fixed
$|z|$ away from the plane `inverts': for example, at fixed
$|z|\approx 2$ kpc, the metallicity {\it increases} with $R$, a trend opposite
to that seen in the midplane.

\begin{figure}
  \hspace{-0.2cm}
  \resizebox{8.5cm}{!}{\includegraphics{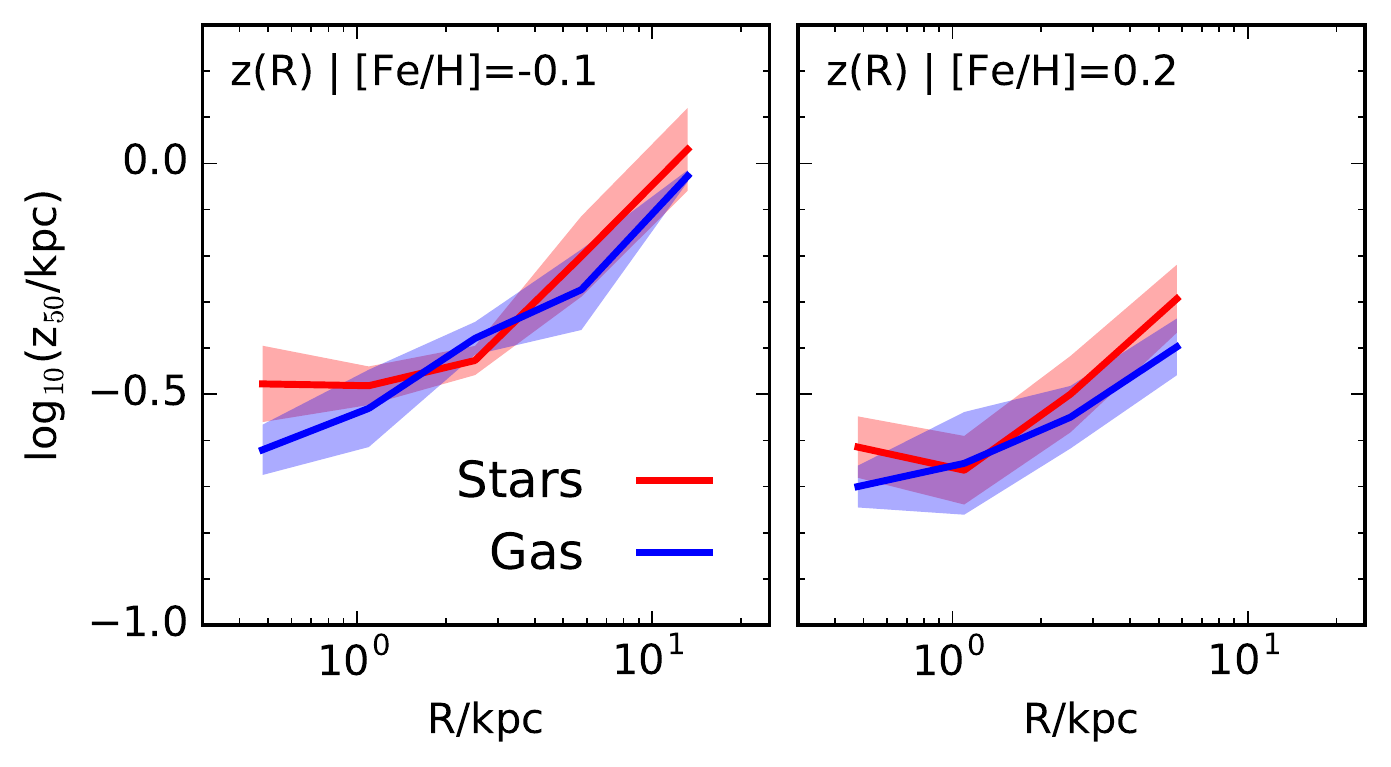}}\\%
  \caption{Red curves indicate the $t=0$  half-mass scaleheight, $z_{50}$, as
    a function of radius $R$, for stars within $0.05$ dex of
    [Fe/H]$=-0.1$ (left) and $+0.2$ (right). These correspond to the
    open circles and squares highlighted in the bottom panels of
    Fig.~\ref{FigGasStarGrad}. Shaded bands indicate the rms scatter
    about the mean. The $z_{50}$ radial dependence of each
    of those populations is in close agreement with the scaleheight of
    the gas at the time when, at each radius $R$, the average
    metallicity matched the selected value (see blue curves). This
    indicates that the `flare' in stars of fixed metallicity is
    inherited from the properties of the gaseous disc at the time of
    their birth and evolves little thereafter.}
  \label{FigMetZGrad}
\end{figure}

\subsection{The origin of radial and vertical gradients}
\label{SecGradOrig}

The origin of the gradients highlighted in the previous subsection may
be elucidated by contrasting the properties of stars at $t=0$ with
those of their parent gaseous disc at the time of their formation. We
do this in Fig.~\ref{FigGasStarGrad}, selecting for analysis star particles
formed between $4$ and $10$ Gyr ago. This period is simpler to analyze
because it is characterized by a
single major accretion episode, which, as discussed in
Sec.~\ref{SecGxAss}, delivered a relatively large amount of
gas to the forming galaxy. The accreted gas quickly assembles into a thick disc
structure that slowly thins down as it forms stars (see; e.g.,
Fig.~\ref{FigGxEvol}).

This evolution is depicted in the left column of
Fig.~\ref{FigGasStarGrad}, where, from top to bottom, we show the
gas surface density profile, $\Sigma(R)$, its half-mass scaleheight
profile, $z_{50}(R)$, and its average metallicity profile,
[Fe/H]$(R)$, coloured as a function of time. These three panels
show the evolution of a self-enriching, slowly-thinning, flared gas disc that
gradually transforms most of its gas into stars. 

Interestingly, as shown in the right column of
Fig.~\ref{FigGasStarGrad}, stars show, at $t=0$, a very similar
structure to the gas in the radial and vertical directions, when
selecting their ages to correspond to the same times chosen for the
gas in the left panels. The clear correspondence between gas and
stars as a function of time/age indicates that the origin of the
stellar gradients lies in the evolving structure of the gaseous
disc. Stars inherit the gas properties at birth and, to first order, preserve
them until $t=0$.

Note that because the gaseous disc is flared at all times, and thins
down as it enriches, stars of fixed metallicity form at different
times at different radii and with different scaleheights. The flare of
stars of fixed metallicity thus reflects the flaring of the gas
disc, modulated by the gradual thinning that occurs at each radius. 

The open circles and squares in Fig.~\ref{FigGasStarGrad}
illustrate this. These symbols track two different metallicities,
[Fe/H]$=-0.1$ and $+0.2$, respectively.  In the bottom-left panel, the
colour underneath each symbol indicates {\it at what lookback time}, as
a function of radius, the average gas metallicity reached each of
those values. In the bottom-right panel, on the other hand, the colours
indicate, as a function of radius, the average {\it age} of star particles
with each of those metallicities, at $t=0$.  The good agreement
between gas and stars demonstrates that most stars have not
migrated far from the radii where they were formed.

The lookback times (for the gas) and ages (for the stars) mentioned in
the preceding paragraph are then used in the middle panels of
Fig.~\ref{FigGasStarGrad} to indicate the half-mass scaleheight of the
gas at each of those times (left panel) or the half-mass scaleheight
of the stars of each of those ages at $t=0$. Again, the close resemblance
between the flare of gas and stars selected in this way indicates that the stellar
gradients originate in the properties of the gaseous disc at the time
of the formation of each coeval stellar population.

One consequence of this resemblance is that, at $t=0$, star particles of fixed
metallicity are flared (see Fig.~\ref{FigMetZGrad}). The steepness of
the flare depends on a combination of several factors: how strong the
gas disc flare is, how quickly the gas self-enriches, and how fast the
disc thins down at various radii.

Had the disc thinned down much more rapidly, for example, so that at
late times the scaleheight in the outskirts was similar to that of the
inner regions at early times, the flare in stars of fixed metallicity
would be much weaker, or might even disappear altogether.  This
interplay between thinning, enrichment, and the flare of the original
gaseous disc explains also why the flaring depends on
metallicity. Taking stars of [Fe/H]$=0.2$ and repeating the exercise,
we find that the flare is less pronounced than that of [Fe/H]$=-0.1$
at $t=0$ (see the right panel of Fig.~\ref{FigMetZGrad}).

\begin{figure}
  \hspace{-0.2cm}
  \resizebox{8.5cm}{!}{\includegraphics{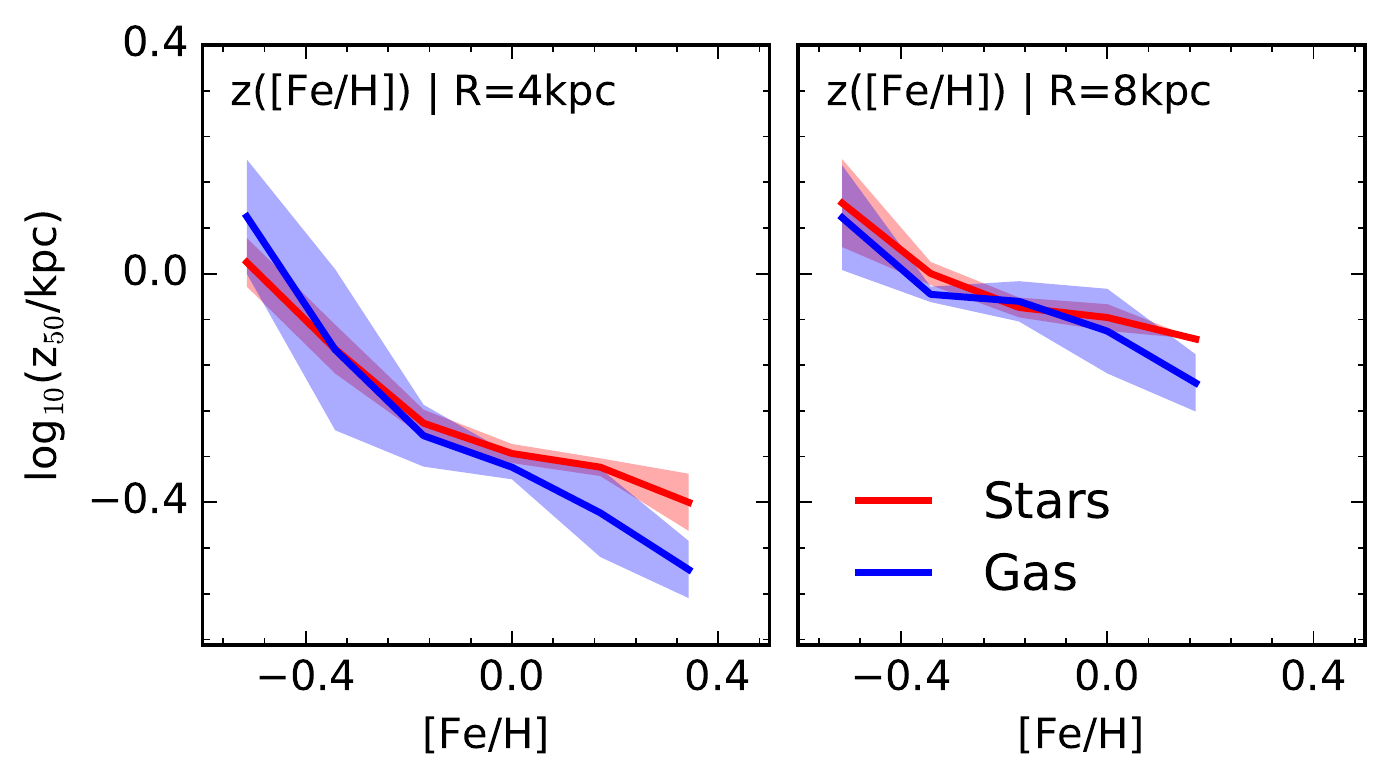}}\\%
  \caption{Half-mass scaleheight, $z_{50}$, as a function of
    metallicity, for stars found at the present time at two selected
    radii; $R=4$ (red curve in left panel) and $8$ kpc (red curve in
    right panel). As in Fig.~\ref{FigMetZGrad}, the blue curves in
    each panel indicate the evolution of the scaleheight of the gas,
    measured at the time when the average metallicity at each radius
    matched the value of [Fe/H] along the $x$-axis. The close agreement
    between gas and stars indicates that the vertical stellar gradient
    is due to the gradual enrichment of a slowly thinning gaseous
    disc. Stars inherit the properties of the gas at birth and
    largely preserve them to the present time.}
  \label{FigMetZGrad2}
\end{figure}

The same exercise also explains the origin of vertical metallicity
gradients at fixed $R$. This is shown in Fig.~\ref{FigMetZGrad2},
where the red curves indicate the half-mass scaleheight of stars as a
function of metallicity, at the present time, and for two different radii, $R=4$ kpc (left)
and $R=8$ kpc (right). The negative gradients (scaleheights decrease
with increasing [Fe/H]) result from the fact that the parent gaseous
disc was gradually thinning down as it enriched. 

This is confirmed by the blue curves in Fig.~\ref{FigMetZGrad2}, which
indicate the half-mass scaleheight of the gas disc {\it at the time}
when its average metallicity at that radius reached the value listed
in the abscissa of Fig.~\ref{FigMetZGrad2}.  In other words, [Fe/H] is
a proxy of time for the gas, and, in the absence of substantial
accretion, tracks the enrichment process of the gaseous disc. The
excellent agreement between the vertical gas evolution and the stellar
gradients at $t=0$ indicate, again, that stars provide, to first
order, a snapshot of the properties of the gas disc at the time of
their formation.

\subsection{Stellar migration after formation}
\label{SecStarDiscEvol}

\begin{figure}
  \hspace{-0.2cm}
  \resizebox{8.5cm}{!}{\includegraphics{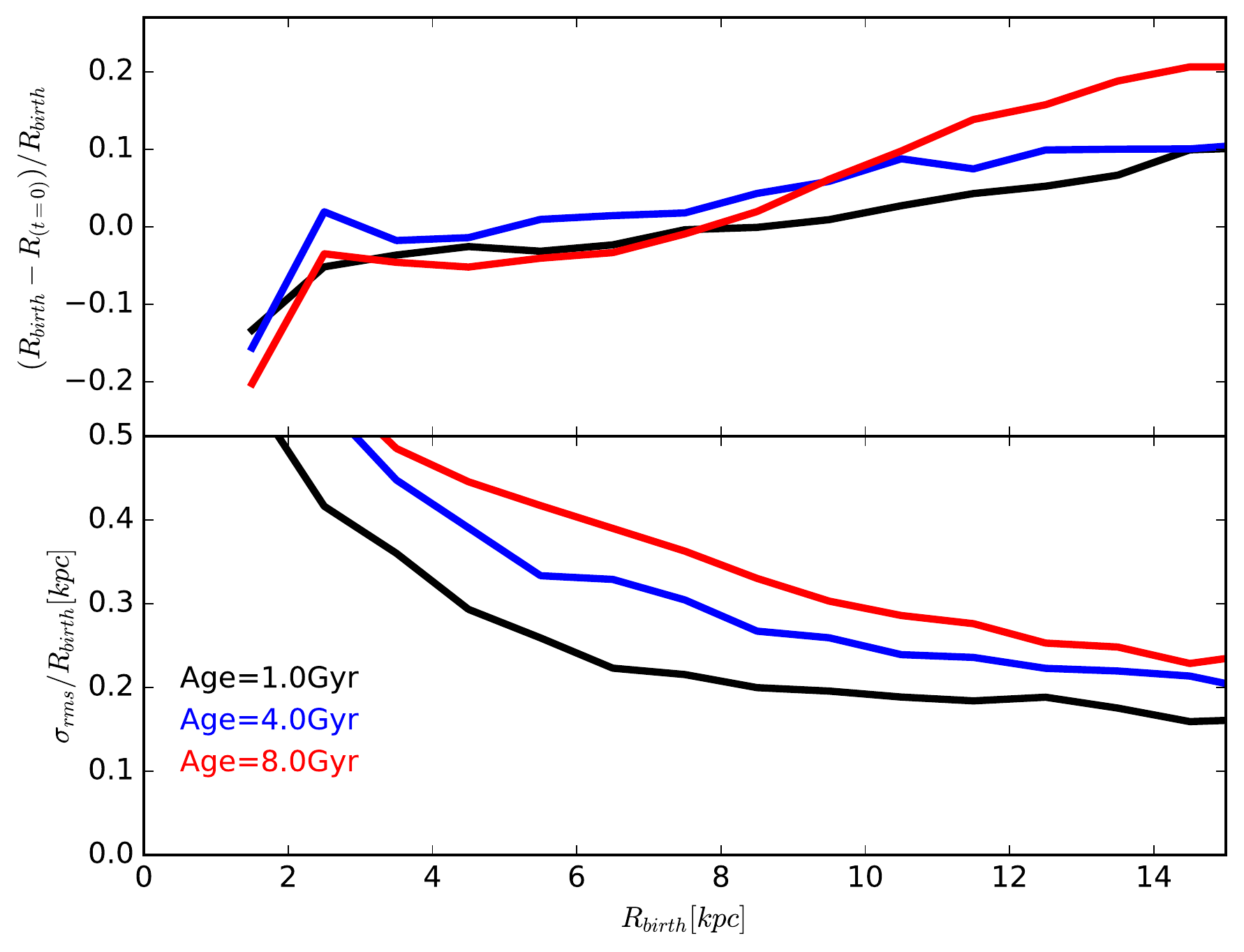}}\\%
  \caption{Radial migration of stars in the simulated disc. For
    particles within $|z|<1$ kpc from the disc, grouped in 2 Gyr-wide
    age bins, we show, as a function of their cylindrical birth
    radii, $R_{\rm birth}$,  their average radii at $t=0$ (top panel)
    and the rms of their radial deviations (bottom panel).}
  \label{FigRadMig}
\end{figure}

The previous discussion suggests that the kinematics of star particles
evolve little after their formation. We show this quantitatively
  in Fig.~\ref{FigRadMig}, where the top panel shows, as a function of
  initial cylindrical birth radius, $R_{\rm birth}$, and for different
  stellar age bins, the average net change in distance from the galaxy
  rotation axis. The net changes are rather small ($<20\%$ over the
  whole radial extent of the disc), indicating that little net
  migration has happened since stars were born. More importantly, the
  rms radial change is also rather small (as seen in the bottom panel
  of Fig.~\ref{FigRadMig}), dropping from $\sim 50\%$ near the centre
  to $\sim 20\%$ in the outskirts of the disc, even for stars as
  old as $8$ Gyr.

\begin{figure}
  \hspace{-0.2cm}
  \resizebox{8.5cm}{!}{\includegraphics{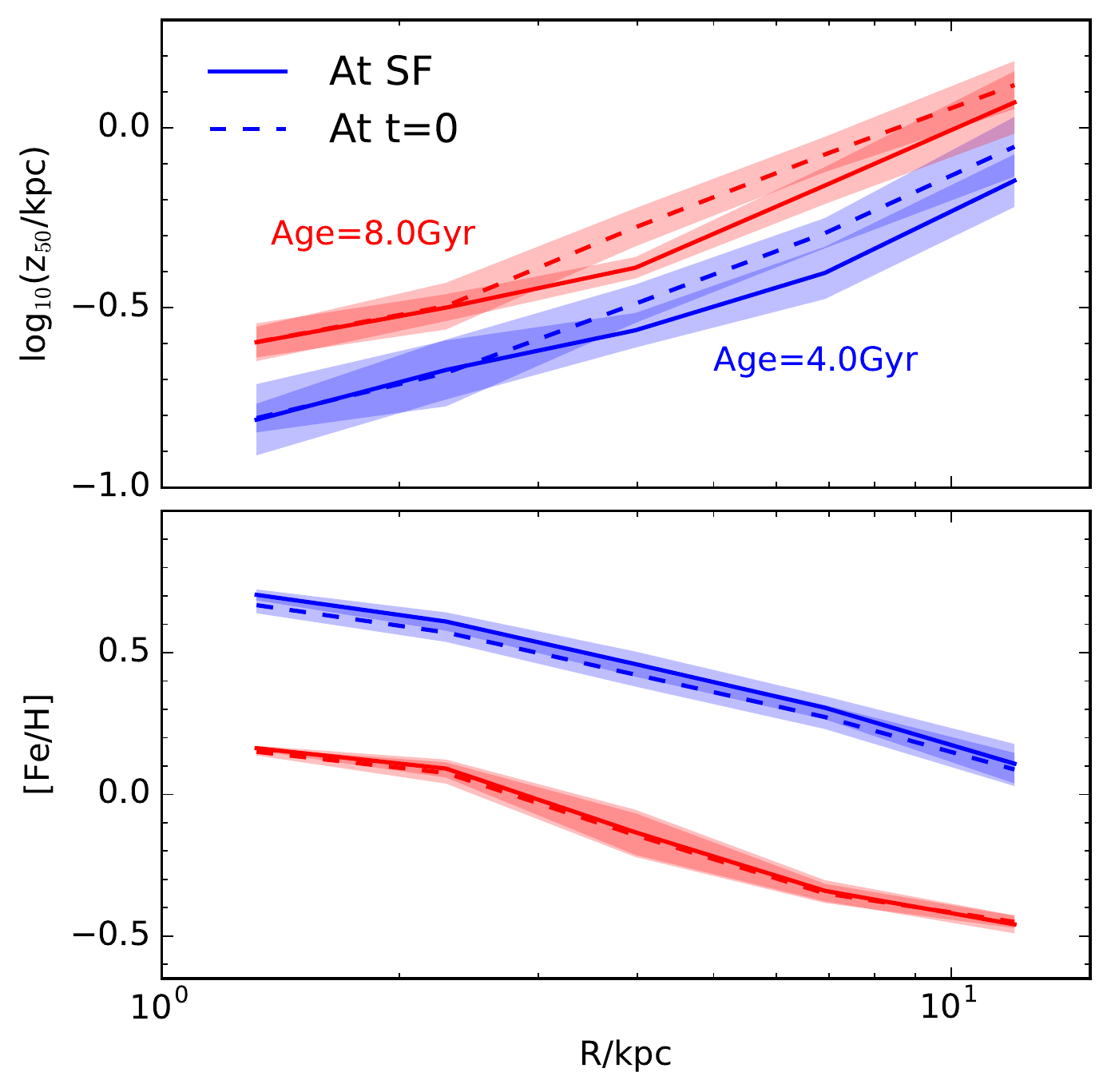}}\\%
  \caption{Radial profiles of the vertical half-mass scaleheight
    (z$_{\rm 50}$; top panel) and metallicity (bottom) of stars shown
    at the time of their formation (solid line) and at the present
    (i.e., lookback time $t=0$; dashed line).  Red and blue correspond to stars with age (at $t=0$) of $4$ and $8$ Gyr respectively. The weak changes with time indicate that the present-day properties of stars were imprinted largely at birth, and evolved little since their formation.}
  \label{FigRGrads}
\end{figure}

The changes induced on radial gradients by migration are consequently
rather small. We demonstrate this explicitly in Fig.~\ref{FigRGrads},
where we show the radial profiles of the vertical scaleheight (top)
and of the metallicity (bottom) of stars formed at two different
lookback times. Dashed lines indicate the properties of such stars at
$t=0$ and solid lines the same at the time of their formation. This
figure shows explicitly that there is little radial mixing of stars
after their formation: the radial metallicity gradients of stars of
given age are essentially the same at $t=0$ as at the time of
formation (bottom panel). There is also little `heating' in the
vertical kinematics of stars after formation: the vertical gradients
are again very similar at the time of formation as at $t=0$ (top
panel).

Since the simulated disc lacks a prominent bar, massive spiral
  arms, or dynamical analogues of giant molecular clouds, it is
  perhaps not unexpected that secular evolutionary processes such as
  radial migration play a minor role in the disc's dynamical
  evolution. Migration in galaxies with stronger spiral patterns
  \citep[and, in particular those where the pattern recurs
  periodically; see, e.g.,][]{Vera-Ciro2014} is expected to be more
  important, and our results thus do not exclude that migration may
  have played an important role in the Milky Way.  This, however, does
  not affect our main result: the radial and vertical trends in the
  simulated disc reflect the conditions of the gaseous disc at birth,
  with secular evolutionary processes playing a minor role.

\subsection{Vertical thinning of the gas disc}
\label{SecVertThin}

The results presented above indicate that star particles of fixed age faithfully
trace the properties of the gaseous disc at the time of their
formation: their vertical and radial gradients are set at birth, and
evolve little thereafter. This elicits two important questions: (i)
what determines the vertical scaleheight of the gas, its outward
flare, and the timescale of its thinning?; and (ii) since, unlike the
gas, stars are not subject to hydrodynamical forces, why do stars
trace the properties of the gas so closely?

The scaleheight of an equilibrium gaseous disc is given by the balance
between pressure forces and the combined vertical compressive forces
of the dark matter halo and the disc \citep[see][for a recent
discussion]{Benitez-Llambay2017}. For an isothermal gas disc with
sound speed $c_s$, embedded in a dark matter halo with circular
velocity profile $V_c(r)$, and surface density profile $\Sigma(R)$, it
is straightforward to show that, in the thin disc approximation, the
vertical compressive force is given by
\begin{equation}
{\partial \Phi \over \partial z}={V_c^2 \over R} {z \over R}
+ 2\pi G \Sigma(z),
\end{equation}
where the first term of the right side indicates the contribution of the dark matter
halo and the second that of the disc, which may also include a stellar
component. ($\Sigma(z)$ is the surface density enclosed between $\pm z$.)

When the halo term dominates, the disc is usually termed ``non-self-gravitating'' (NSG)
and its characteristic scaleheight is given by
\begin{equation} 
z_{\rm NSG}={c_s \over V_c(R)} R.  
\label{EqZNSG}
\end{equation} 
On the other hand, when the disc term dominates the vertical force the disc
is ``self-gravitating'' (SG) and its characteristic scaleheight is given by
\begin{equation}
z_{\rm SG}={c_s^2 \over G\Sigma(R)},
\label{EqZSG}
\end{equation}
where $\Sigma(R)$ is total surface density integrated over all $z$.
Because the vertical density law differs when the disc is SG or NSG
the characteristic scaleheight values given by Eqs.~\ref{EqZNSG} and
\ref{EqZSG} are not direct measures of the half-mass scaleheights,
$z_{50}$, which is what we actually measure in the
simulations. The proportionality factors, however, may be easily
computed in each case and may be consulted in
\citet{Benitez-Llambay2017}.

Note that, in general, we expect isothermal exponential gas discs with
flat circular velocity curves to `flare' outwards, since, in that
case, the thickness would be $\propto R$ if non-self-gravitating, and
$\propto \Sigma^{-1}$ if self-gravitating.

\begin{figure}
  \hspace{-0.2cm}
  \resizebox{8.5cm}{!}{\includegraphics{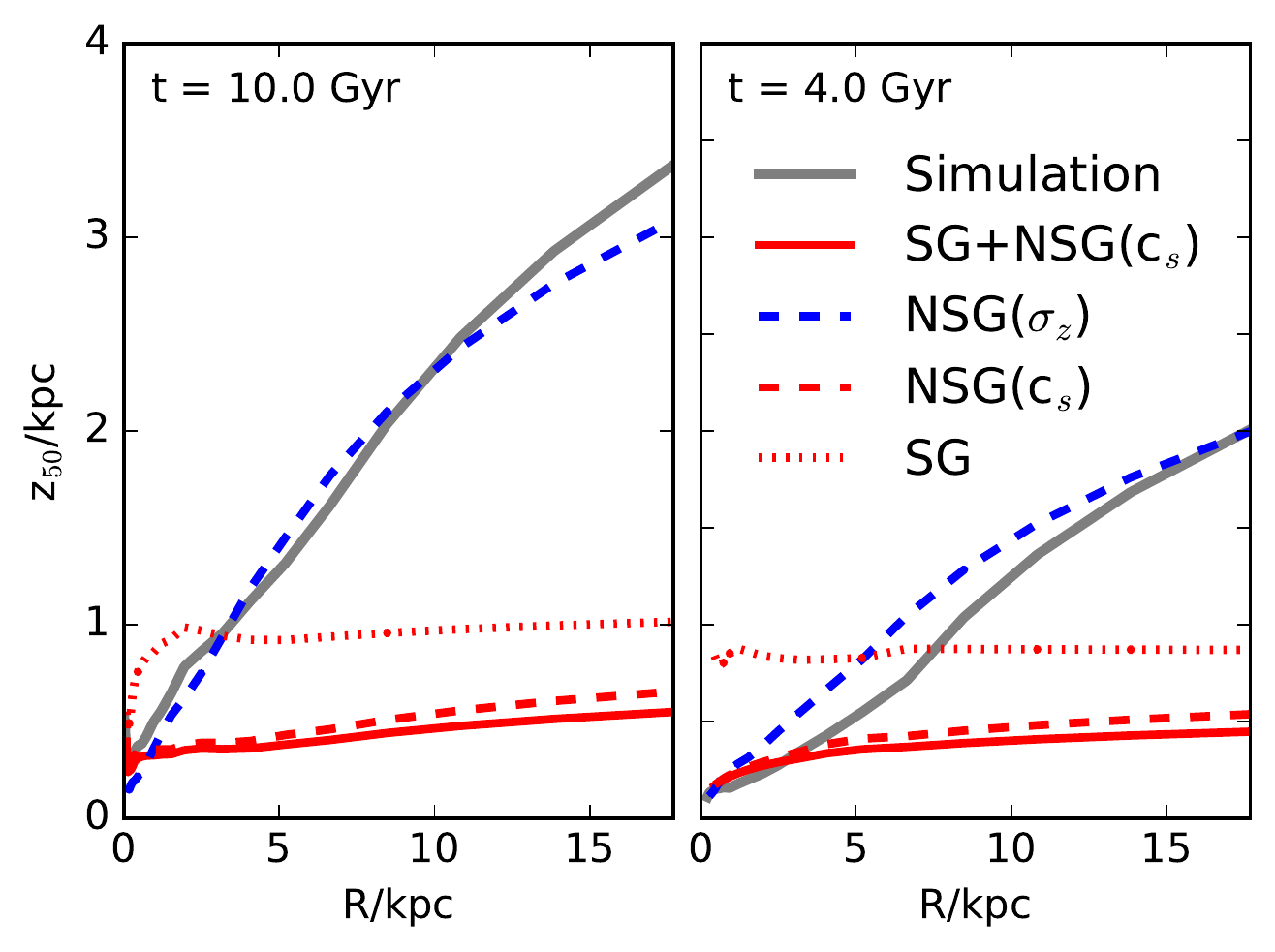}}\\%
  \caption{The (strongly flared) vertical half-mass scaleheight
    profile of the gas at $t=10$ Gyr and $t=4$ Gyr (thick grey solid
    line). Expected scaleheights assuming vertical hydrostatic
    equilibrium with thermal pressure are indicated for the case of a
    self-gravitating disc (SG, red dotted) and non-self-gravitating
    disc (NSG, red dashed). These are generally much smaller than the
    actual thickness of the simulated disc, which is well matched by
    the NSG solution but where the `pressure' support is given by
    the vertical velocity dispersion of the gas particles ($\sigma_z$)
    rather than by the midplane thermal sound speed ($c_s$). The disc
    is largely kept thick by the feedback-induced bulk motions of the
    gas. (See text for a full discussion.)}
  \label{FigRZ50}
\end{figure}

We compare in Fig.~\ref{FigRZ50} the above expectation with the gas
disc half-mass scaleheights measured at two different lookback times,
$\approx 10$ and $\approx 4$ Gyr ago. The thick grey curve shows the result
of the simulation, whereas the SG and NSG scaleheights are shown with
dotted and dashed red curves, respectively. We use the midplane sound
speed to account for the density-dependent EAGLE equation of state; as
a result, the effective sound speed in Eq.~\ref{EqZNSG} depends
(weakly) on $R$. The solid red line is the actual expected value when
considering the disc and halo combined vertical forces.  Clearly the
simulated disc is {\it much thicker} than expected given these
considerations, at essentially all radii.

Indeed, what sets the thickness of the disc is actually the balance
between the vertical forces and the `pressure' provided by random bulk
motions in the gas. The blue dashed curves in Fig.~\ref{FigRZ50} show
again the NSG solution, but replacing the sound speed, $c_s$, with
the vertical velocity dispersion of the gas particles, $\sigma_z$, in
Eq.~\ref{EqZNSG}. The agreement between this curve and the simulation
results indicates that the gas disc is kept thick by the vertical
random motions of the gas, and, in particular, of the star forming gas
clouds that may be clearly seen in the bottom panels of
Fig.~\ref{FigGxEvol}, especially at early times.

These random motions should be quelled in a couple of dynamical
timescales, but, as seen in the right panel of
Fig.~\ref{FigRZ50}, they still dominate in the outskirts of the disc
after $6$ Gyr of evolution, or more than $\approx 20$ circular orbit
periods at $R=10$ kpc. (Orbital times as a function of radius, as well
as circular velocity profiles of the simulated galaxy are provided in
the Appendix.) The gas disc is actually kept thick by the feedback
energy provided by evolving stars---this is deposited as thermal
energy in the disc, where it is able to blow bubbles and to push gas
outside the disc quite efficiently, especially when star formation
rates are high.

\begin{figure}
  \hspace{-0.2cm}
  \resizebox{8.5cm}{!}{\includegraphics{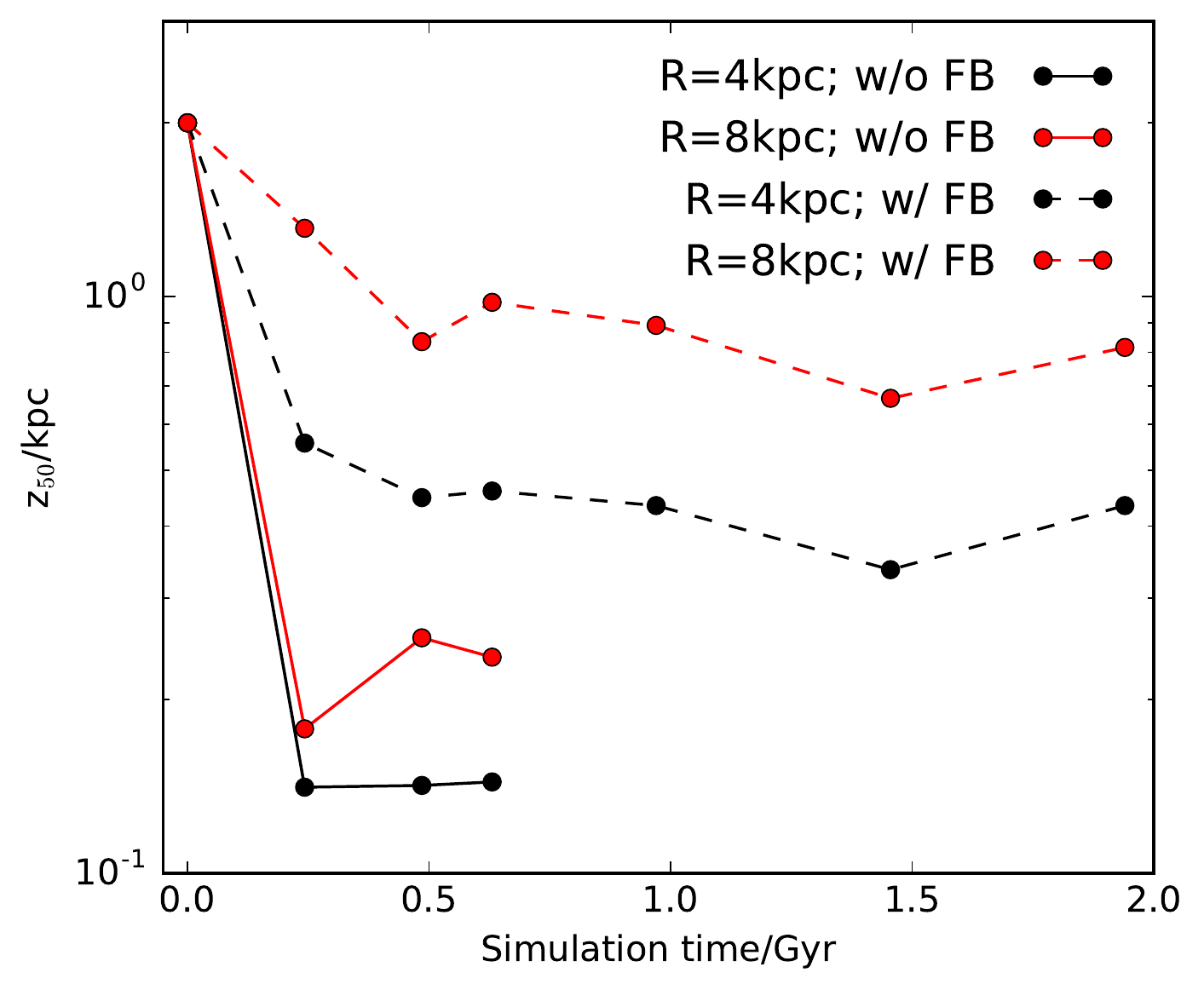}}\\%
  \caption{Scaleheight evolution, at two different radii,  of an idealized gas disc simulation
    with (dashed lines) and without (solid lines) the effects of star
    formation and feedback. Note that the disc is much thicker when
    including star formation, demonstrating the key role of feedback
    in stirring up the vertical motions that keep the gas disc
    thick. See text for details on the idealized simulation.}
  \label{FigAle}
\end{figure}

We show this in Fig.~\ref{FigAle}, where we plot the half-mass
scaleheight evolution of a gas disc formed in an idealized simulation
which evolved a system tailored to match approximately the mass, size,
and angular momentum of the gaseous disc at $t=10$ Gyr (see
Fig.~\ref{FigGasStarGrad}). In this simulation, gas is allowed to cool
and collapse from a $\approx 2$ kpc-thick rotating `slab' embedded in a
spherical Navarro-Frenk-White \citep[NFW,][]{Navarro1996,Navarro1997}
dark matter halo potential with parameters chosen to match the
circular velocity profile of the galaxy at that time
(Fig.~\ref{FigRVcirc}).

The solid lines in Fig.~\ref{FigAle} indicate the evolution of the
half-mass scaleheight at two different radii ($R=4$ and $8$ kpc), when
the system is evolved in the absence of star formation or feedback. As
expected, the disc quickly collapses vertically to a
pressure-supported disc roughly $\approx 200$ pc thick. The simulation is
ended after just $\approx 700$ Myr because the disc breaks into massive
self-bound clumps, making estimates of its thickness unreliable.  When
star formation is included, however, the gaseous disc settles into a
much thicker structure which is supported by the gas bulk motions
induced by the feedback energy released by evolving stars. (See dashed
lines in Fig.~\ref{FigAle}.) 

Interestingly, the idealized disc with feedback is still thinner than
the APOSTLE galaxy (compare with the left panel of
Fig.~\ref{FigRZ50}), suggesting that there may be an additional source
of vertical support in the APOSTLE simulation, possibly related to
bulk motions induced by the continuous gas accretion, which is not
included in the idealized runs. In any case, Fig.~\ref{FigAle} shows
clearly that the thickness of our simulated disc is largely set and
controlled by feedback-induced vertical motions in the gas, whose
effective pressure far exceeds the thermal support.

Feedback is thus a critical ingredient for understanding the origin of
vertical gradients in our simulations, connecting the star formation
history at each radius and its consequent enrichment with the evolving
thickness of the star forming disc.  In this scenario, the disc thins
down as the gaseous disc is depleted, star formation abates, and
feedback (and, possibly, accretion) heating becomes less effective. A
strong prediction is then that the star formation, enrichment, and thinning
timescales should all be linked at all radii, as it may have already
been observed in the inner Galaxy \citep[see;
e.g.,][]{Freudenburg2016}.

The same scenario explains why the vertical structure of stars tracks
so closely that of the gas at the time of their formation. This is
because the star forming gas is best described as a collection of
dense clouds with appreciable bulk velocities, rather than as a fluid
in vertical hydrostatic equilibrium supported by thermal
pressure. Indeed, had star particles been born of gas in thermal hydrostatic
equilibrium the stellar disc would be much thinner than its parent gas
disc due to the loss of vertical pressure
\citep[see][]{Benitez-Llambay2017}. In our simulations, however, star particles
inherit the bulk motions of the gas clouds from which they are
born. Since these motions are dominant over thermal pressure forces
there is little difference between gas clouds and newly formed stars,
explaining why stars trace faithfully the properties of the gas at the
time of their formation.

We end this discussion with a caveat by noting that, when feedback is
included, the gaseous disc thickness exceeds that of the thin disc of
the Milky Way (see Fig.~\ref{FigAle}). This elicits questions as to
whether a more realistic star formation/feedback model would yield
radial and, in particular, vertical trends similar to those we report
here. A definitive answer to this question will need to wait until our
modeling improves to match the observed disc thickness of Miky
Way-like spirals. However, supporting evidence for our conclusions may
be gleaned from simulations that yield much thinner discs than ours
but, qualitatively, the same trends we report here \citep[see,
e.g.,][and references therein]{Ma2017}. We are planning to revisit
this issue in future work.

\section{A model for the origin of metallicity gradients}
\label{SecModel}

The scenario discussed above, where gradients result from the
intertwined evolution of the gaseous disc's star formation,
enrichment, and vertical structure, may be encapsulated in a simple
model that can be used to interpret the origin of various chemical and
kinematical trends in the Galaxy. The model assumes that stars inherit
the properties of the gas at the time of formation and preserve them
to the present. It also requires, at a minimum, a characterization of
the disc surface density profile, $\Sigma(R)$, its thickness profile,
$z_{50}(R)$, and its average metallicity profile,
$\langle$[Fe/H]$\rangle (R)$, as a function of time.

\subsection{Gas disc evolution parametrisation}
\label{SecEvolDiscParam}

The evolution of the vertical structure of the APOSTLE gaseous disc 
may be approximated simply by
\begin{equation}
  z_{50}(R)=z_{50,0} \ e^{{R}/{R_z}} \ e^{(t/\tau_z)^2},
\label{Eqz50Evol}
\end{equation}
where $t$ is lookback time, $z_{50,0}=z_{50}(R=0,t=0)$ is the central
half-mass disc scaleheight at the present time, $R_z$ is a
characteristic flaring radius, and we have assumed that the thinning
timescale of the disc, $\tau_z$, is independent of $R$.

A similar description may be adopted for the evolution of the
metallicity gradient. We assume that the average metallicity profile
is given by $Z(R,t)=\langle$[Fe/H]$\rangle(R,t)$, and that it may be
approximated by
\begin{equation}
{Z}(R,t)= {Z_{0}}\,  [1+\alpha_0 (R/Z_0)]\, \ e^{-(t/\tau_Z)^2},
\label{EqZEvol}
\end{equation}
where $Z_0=Z(R=0,t=0)$ is the central average gas metallicity at the
present time, $\alpha_0=(dZ/dR)_{R=0}$, and we have 
assumed that the enrichment timescale, $\tau_Z$, is independent of
$R$.

\begin{figure}
  \hspace{-0.2cm}
  \resizebox{8.5cm}{!}{\includegraphics{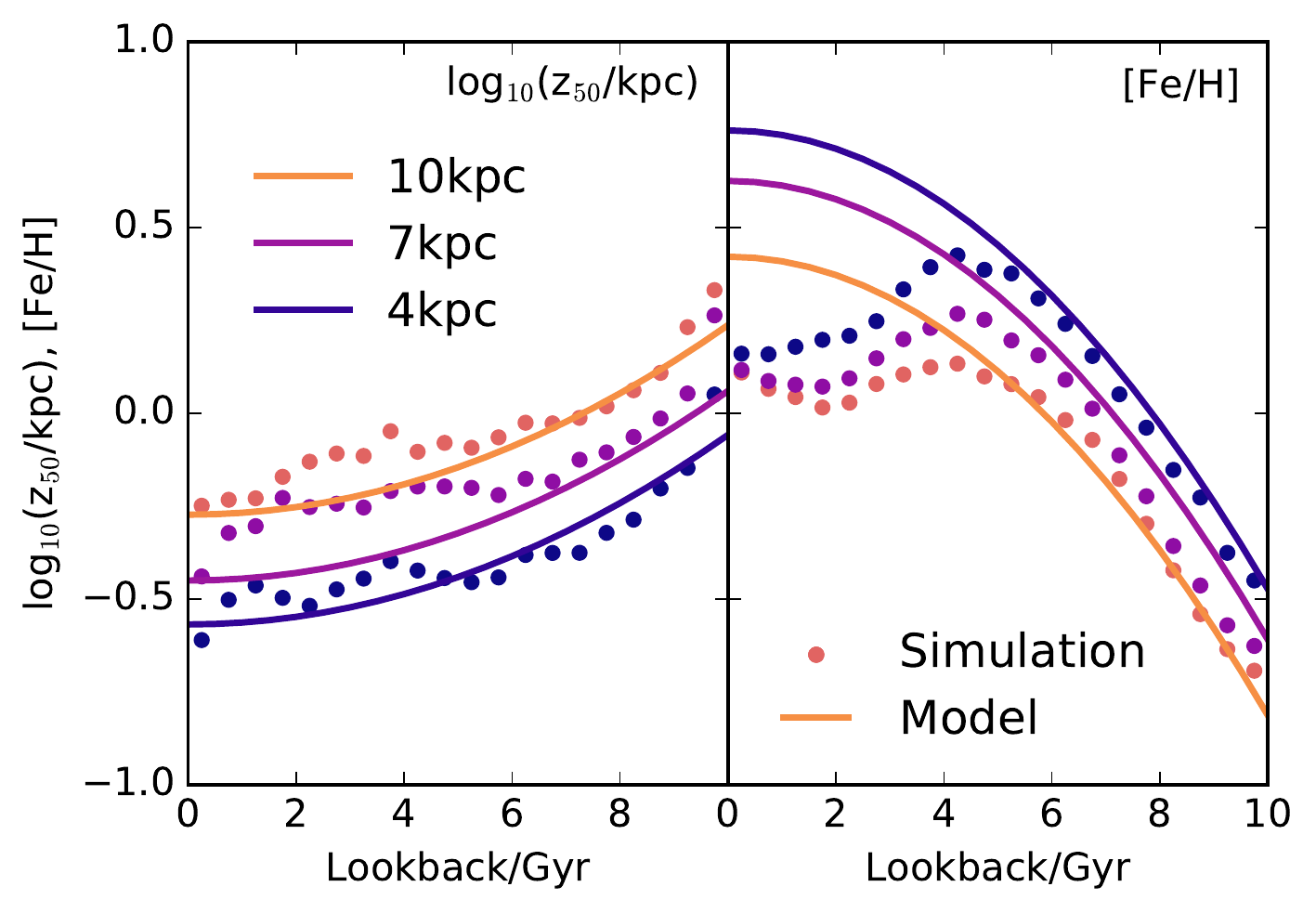}}\\%
  \caption{Evolution with lookback time of the gas scaleheight (left) and
    average metallicity (right) at three different radii: $R=10$, $7$,
    and $4$ kpc. The solid circles indicate the result of the
    simulation, the thick lines are fits using the parametrisation
    given in Eq.~\ref{Eqz50Evol} and Eq.~\ref{EqZEvol}.}
  \label{FigTZ50}
\end{figure}

We compare these parametrisations with the evolution of the simulated
disc in Fig.~\ref{FigTZ50}, for three different radii. The thinning
and enrichment timescales that best approximate the simulated disc are
$\tau_z=14$ Gyr and $\tau_Z=9$ Gyr, respectively. This figure shows
that the functional forms adopted above are adequate, at least for
$t>4$ Gyr. At more recent lookback times the disc accretes a
substantial amount of metal-poor gas (see middle panel of
Fig.~\ref{FigMassEvol}), lowering the average metallicities at all
radii. This non-monotonic behaviour cannot be reproduced by our simple
formulation.

\subsection{Application to the Milky Way}
\label{SecModelIll}

Given the assumptions of the parametrisation adopted above, the
radial and vertical gradients that result will be largely set by the
flaring profile and metallicity gradient adopted for the gaseous disc
at $t=0$, as well as by the ratio between the thinning timescale,
$\tau_z$, and the enrichment timescale, $\tau_Z$. Some intuition may be
gained from considering hypothetical cases when one timescale is much
longer than the other.

For example, if enrichment proceeded much faster than thinning, then
all populations, regardless of metallicity, would have the same
flaring profile (that of the gaseous disc), and there would be no
vertical metallicity gradient at fixed $R$. If, on the other hand,
thinning proceeded much faster than enrichment, then we would expect
strong vertical gradients to develop at fixed $R$. In this case, the
flaring profile of stars of fixed metallicity would depend in detail
on the gas disc flare, as well as on the ratio of thinning
and enrichment timescales. Depending on these parameters, stars of given $Z$
might show no flare even though the gaseous disc is always flared.

We illustrate this next with a simple application to the Milky Way. We
  emphasize that this is {\it not} meant to be quantitative model of
  the chemical evolution of the Milky Way disc(s) but rather an
  illustration of how the results discussed in the previous
  subsections may be used to interpret the data. 

\begin{figure}
  \hspace{-0.2cm}
  \resizebox{8.5cm}{!}{\includegraphics{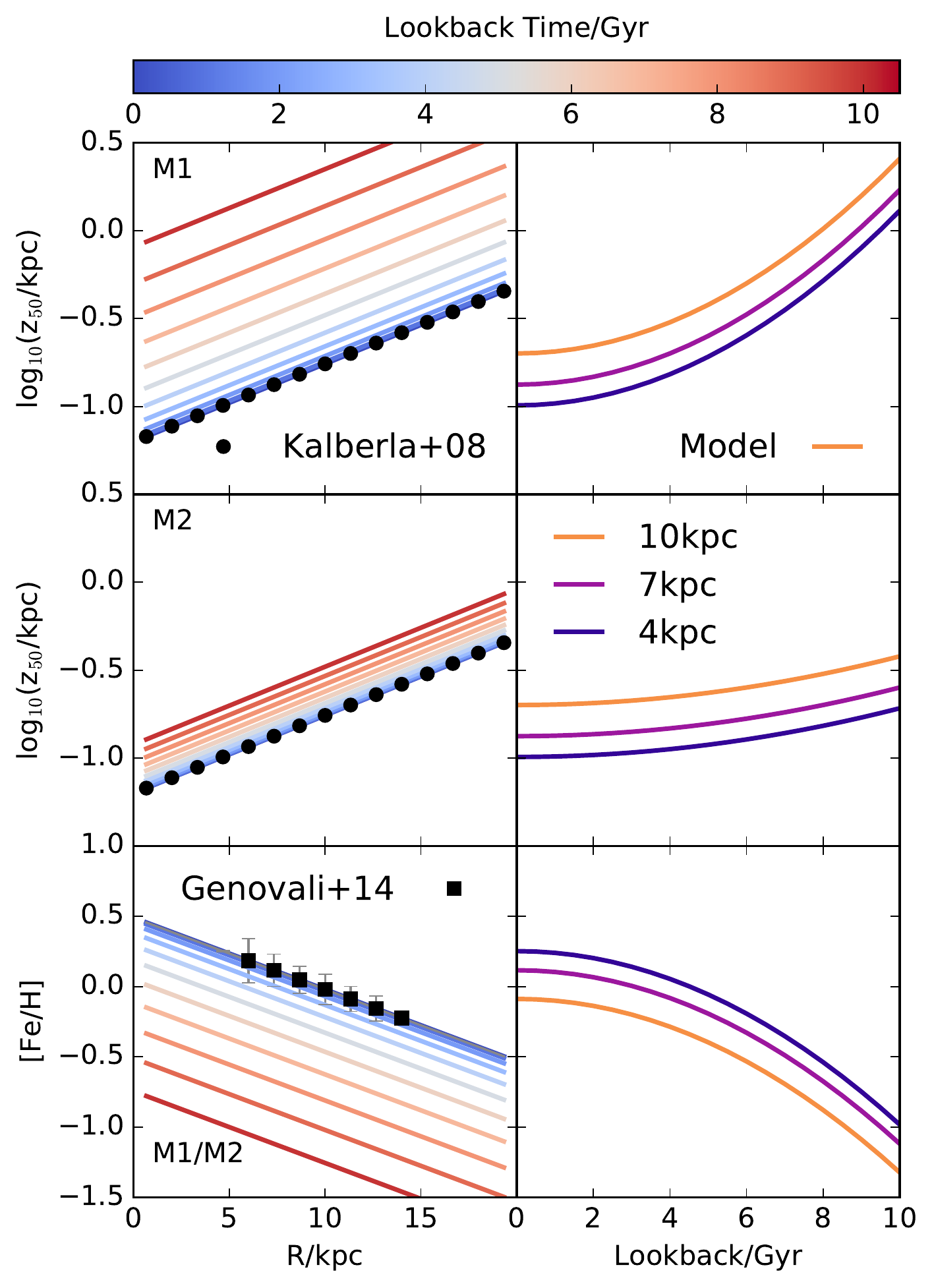}}\\%
  \caption{ {\it Left column:} Top and middle panels illustrate the
    evolution of the gas half-mass scaleheight profile, $z_{50}(R)$,
    assumed in model 1 (M1) and model 2 (M2), respectively.  Bottom
    panel is analogous, but for the average metallicity profile. The
    shapes of the final profiles have been chosen to match the
    exponential fit of \citet{Kalberla2008} (top left) and the Cepheid
    metallicity gradient  of \citet{Genovali2014}. {\it Right column:}
    Scaleheight evolution at three different radii; $R=10$, $7$ and
    $4$ kpc, for M1 (top) and M2 (middle). Bottom panel shows the
    evolution of the average metallicity at the same radii, which is
    the same for both models.}
  \label{FigRModel}
\end{figure}

The model requires the present-day gas MW density/metallicity
profiles. We show these in the left panels of Fig.~\ref{FigRModel}:
the top two show the vertical gas scaleheight profile taken from the
fits to 21-cm data presented by \citet{Kalberla2008}. The gas disc is clearly flared,
and can be well approximated, at $t=0$, by Eq.~\ref{Eqz50Evol}
($z_{50,0}=0.06$ kpc and $R_z=9.8$ kpc), as shown by the bue lines.  The
bottom-left panel shows the Cepheid metallicity profile from
\citet{Genovali2014}, which we shall take to represent the present-day
radial metallicity profile of the Galaxy midplane, and which we
approximate with $Z_0=0.5$ and $\alpha_0=-0.06$ dex/kpc. This gradient
can also be approximated by the parametrisation adopted in
Eq.~\ref{EqZEvol}, as shown by the blue lines in that
panel. 

The only remaining parameters are $\tau_z$ and $\tau_Z$. We choose two
cases, again mainly for illustration. Model 1 (M1) assumes
$\tau_z\approx \tau_Z$; i.e., $\tau_z=9.5$ Gyr, and $\tau_Z=9.0$
Gyr. Model 2 (M2), on the other hand, assumes that thinning proceeds
on a much longer timescale than enrichment; i.e., $\tau_z=19$ Gyr, and
$\tau_Z=9.0$ Gyr. Note that the two models differ only in the thinning
timescale.

The evolution of both thickness and {\it average} metallicity is shown
for three different radii; $R=10$, $7$, and $4$ kpc, in the right
panels of Fig.~\ref{FigRModel}. In M1 the disc thins down quickly from
a much thicker earlier configuration to its final state, whereas it
evolves much more gradually in M2. The metallicity evolution is
identical in both models, which assume that the disc has enriched at
all radii by more than one dex\footnote{We note that the latter
  result differs from the solar neighbourhood, where the
  age-metallicity relation has large scatter and is relatively flat in
  the age range $0$-$8$ Gyr \citep{Nordstrom2004,Haywood2013}. We
  emphasize again that our simulated galaxy is not meant to be a model
  for the Milky Way, but rather a basic illustration of how our
  results may be applied to gain insight into the Milky Way
  evolution.} in the past $\approx 10$ Gyr.

\begin{figure}
  \hspace{-0.2cm}
  \resizebox{8.5cm}{!}{\includegraphics{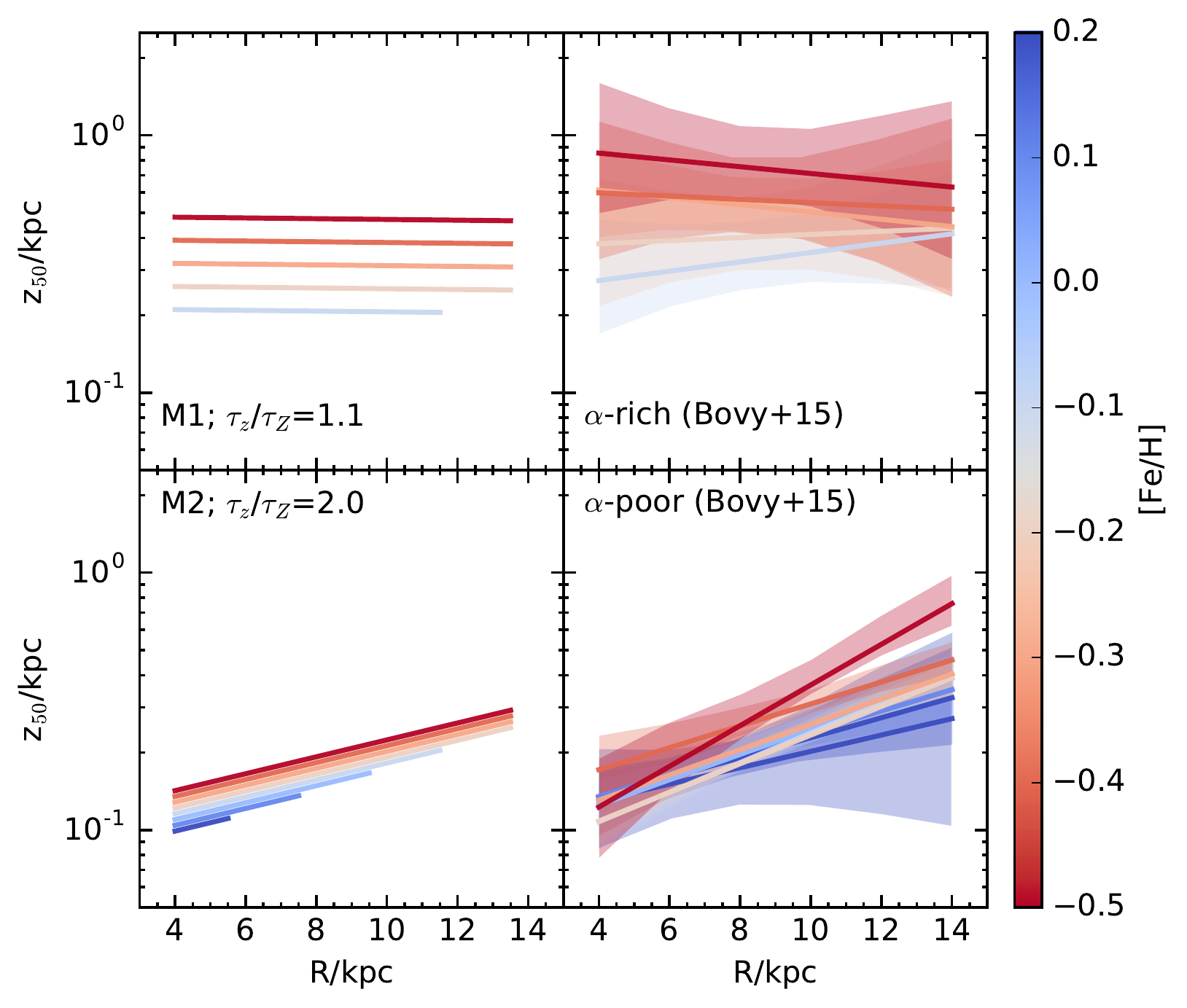}}\\%
  \caption{{\it Left column:} Scaleheight profiles of stars of
    fixed-metallicity formed out of gas discs evolving as in model
    1(top) or model 2 (bottom). Stars are assumed to inherit and
    preserve the properties of the gaseous disc at birth. {\it Right
      column:} A summary of the scaleheight profiles for stars of
    fixed metallicity, taken from \citet{Bovy2016}. Top panel shows
    that fixed-metallicity stars in the $\alpha$-rich (`thick') disc
    show no obvious flares. Stars of the $\alpha$-poor (`thin')
    component, on the other hand, flare outward. Either of these
    trends can be reproduced by models M1 and M2, by adjusting a
    single parameter, the ratio of enrichment-to-thinning timescale,
    $\tau_Z/\tau_z$. See text for a detailed discussion.  }
  \label{FigBovyModel}
\end{figure}

The effect of these two different evolutionary patterns on the flaring
profile of stars of fixed metallicity is shown in the left panels
of Fig.~\ref{FigBovyModel}. In the case of Model 1 the enrichment and
thinning timescales are comparable and, for the flaring profile of
\citet{Kalberla2008}, leads to populations of stars that show {\it no flare}
at fixed metallicity.

Model 2, on the other hand, leads to well-defined flares in
populations of fixed metallicity, as shown in the bottom-left panel of
Fig.~\ref{FigBovyModel}. Interestingly, both trends are seen in the
Milky Way disc, where there is compelling evidence for the presence of
two chemically-distinct populations \citep[][]{Bovy2016}: an
$\alpha$-rich disc (the traditional `thick disc') with no flares
(like M1), and an $\alpha$-poor (`thin') disc whose subpopulations
clearly flare outwards (like M2). (See right panels of
Fig.~\ref{FigBovyModel}.)

Although the presence of these flares has been taken as evidence
  for the effects of radial migration, our results argue that such
  flares might also result from the gradual thinning of the gaseous
  disc. This should not be viewed as implying that radial migration
  does not occur in the Milky Way; only that the mere presence of a flare does not
  guarantee that it has been caused by radial migration, and that
  other explanations should also be considered \citep[see; e.g.,][for
  a more comprehensive discussion of the role of radial migration on
  disc
  flaring]{Schoenrich2009,Loebman2011,Minchev2012,Kubryk2013,Roskar2013,Vera-Ciro2014,Grand2015,Vera-Ciro2016,Kawata2017}.

\section{Summary and Discussion}
\label{SecConc}

We use a cosmological simulation from the APOSTLE project to study the
origin of radial and vertical gradients in the age and metallicity of
disc stars. We focus our analysis on one particular galaxy selected
because of its overall resemblance to the Milky Way: most stars at
lookback time 
$t=0$ are in a well defined, coplanar, centrifugally-supported disc
component; it has formed stars throughout its history; it has had no
recent major mergers; has no major morphological peculiarities; has
had a quiet recent merger history; and has formed most of its stars
in-situ.

Some radial and vertical trends in the simulated galaxy resemble
those of the Milky Way. In particular, average stellar ages and metallicities
decrease with increasing radius in the disc midplane; at fixed $R$,
metallicity decreases and age increases with increasing $|z|$; and the
disc exhibits a pronounced `flare' (i.e., the scaleheight of stars
of fixed age or metallicity increases monotonically outwards). 

We trace the origin of these gradients to the properties of the parent
gaseous disc, which first assembles into a thick, slowly-rotating, flared
structure and then gradually thins
down, settles, and cools kinematically as gas turns into stars and enriches itself in
the process. The disc is denser near the centre, and therefore 
enrichment and thinning proceed more rapidly there than in
the outskirts.

Stars inherit the properties of the gas disc at the time of their
formation, and evolve little thereafter. In other words, the resulting
gradients are predominantly imprinted at birth, and are not the result of secular
evolutionary processes such as radial migration or disc
instabilities.

The similarity in the vertical structure of stars at late times and
that of the gas at the time of their formation results because the
gaseous disc is not in thermal hydrostatic equilibrium. Rather, its
vertical structure is largely set by the bulk motions of star forming
gas clouds, which are in turn induced and sustained by the feedback
energy of evolving stars and, possibly, energy injected by continuous
gas accretion.  As a result, the timescales of star
formation, enrichment, and equilibration are all intertwined, and
leave behind clear radial and vertical gradients in the gaseous disc
and its descendent stars.

Our results suggest that some trends that are often ascribed to
  secular evolutionary processes may actually be the result of the
  gradual equilibration process of a star forming, self-enriching
  disc. This should not be taken to imply that secular evolution plays
  no role in the establishment of such gradients in the case of the
  Milky Way, where the bar and spiral patterns, which are lacking in
  our simulated disc) are expected to lead to some radial and vertical
  redistribution of stars after formation.
 Rather, our results simply
demonstrate that such gradients, per se, should not be regarded as
{\it necessarily} due to secular evolution.

Our simulation is the latest of a number of cosmological
simulations of disc galaxy formation that argue that the conditions
`at birth' play a critical role in our understanding of the origin
of Galactic gradients \citep[see; e.g.,][and references
therein]{Brook2012b,Bird2013,Stinson2013,Grand2015,Miranda2016,Ma2017,Grand2017}. Although
the simulations differ in their details, the qualitative scenario they
recount is common to all: a feedback-thickened, star-forming, flared
gaseous disc that gradually turns itself into stars as it equilibrates
and settles down, results in a disc galaxy that resembles the Milky Way
in a surprising number of ways. These simulations also show that
internal evolutionary processes (such as migration and vertical
thickening) do matter, but are neither responsible for the main structural
properties of the galaxies, nor for the origin of their gradients.

This scenario offers a theoretical template that may be used to
interpret observations, such as those that will result from ongoing
and upcoming surveys of the Milky Way. Its crucial prediction is that
locally the timescales of star formation, enrichment, equilibration,
and thinning should all be connected through a simple physical model
of feedback and accretion. The relations between these timescales and their
observational consequences need to be spelled out in more detail by
future work to yield falsifiable predictions that may be used to
assess the validity and general applicability of this scenario. 

\section{Acknowledgements}

The research was supported in part by the Science and Technology
Facilities Council Consolidated Grant (ST/F001166/1), and the European
Research Council under the European Union's Seventh Framework
Programme (FP7/2007-2013)/ERC Grant agreement
278594-GasAroundGalaxies. CSF acknowledges ERC Advanced Grant 267291
COSMIWAY. This work used the DiRAC Data Centric system at Durham
University, operated by the Institute for Computational Cosmology on
behalf of the STFC DiRAC HPC Facility (www.dirac.ac.uk). The DiRAC
system was funded by BIS National E-infrastructure capital grant
ST/K00042X/1, STFC capital grants ST/H008519/1 and ST/K00087X/1, STFC
DiRAC Operations grant ST/K003267/1 and Durham University. DiRAC is
part of the National E-Infrastructure. This research has made use of
NASA's Astrophysics Data System.

\section{Appendix}

\begin{figure}
  \hspace{-0.2cm}
  \resizebox{8.5cm}{!}{\includegraphics{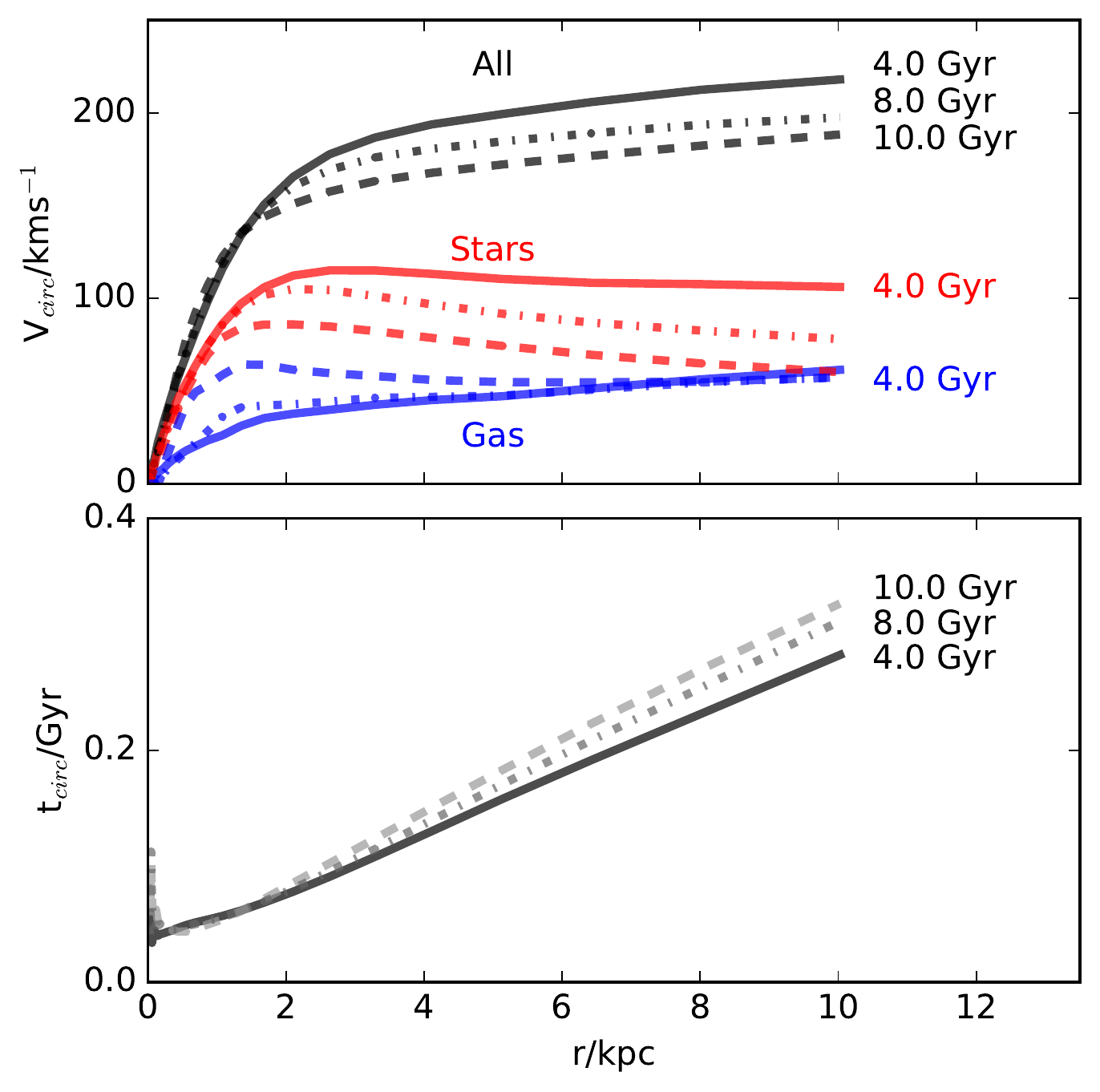}}\\%
  \caption{ Top panel shows circular velocity profiles (assuming
    spherical symmetry), $V_{\rm circ}(R)=(GM(R)/R)^{1/2}$, at $t=0$
    and selected lookback times ($7.5$ and $9.5$ Gyr ago).  Bottom
    panel shows the circular orbital time as a function of spherical
    radius, assuming $t_{\rm circ}=2\pi r/V_{\rm circ}$, at the same
    lookback times.}
  \label{FigRVcirc}
\end{figure}

\begin{figure}
  \hspace{-0.2cm}
  \resizebox{8.5cm}{!}{\includegraphics{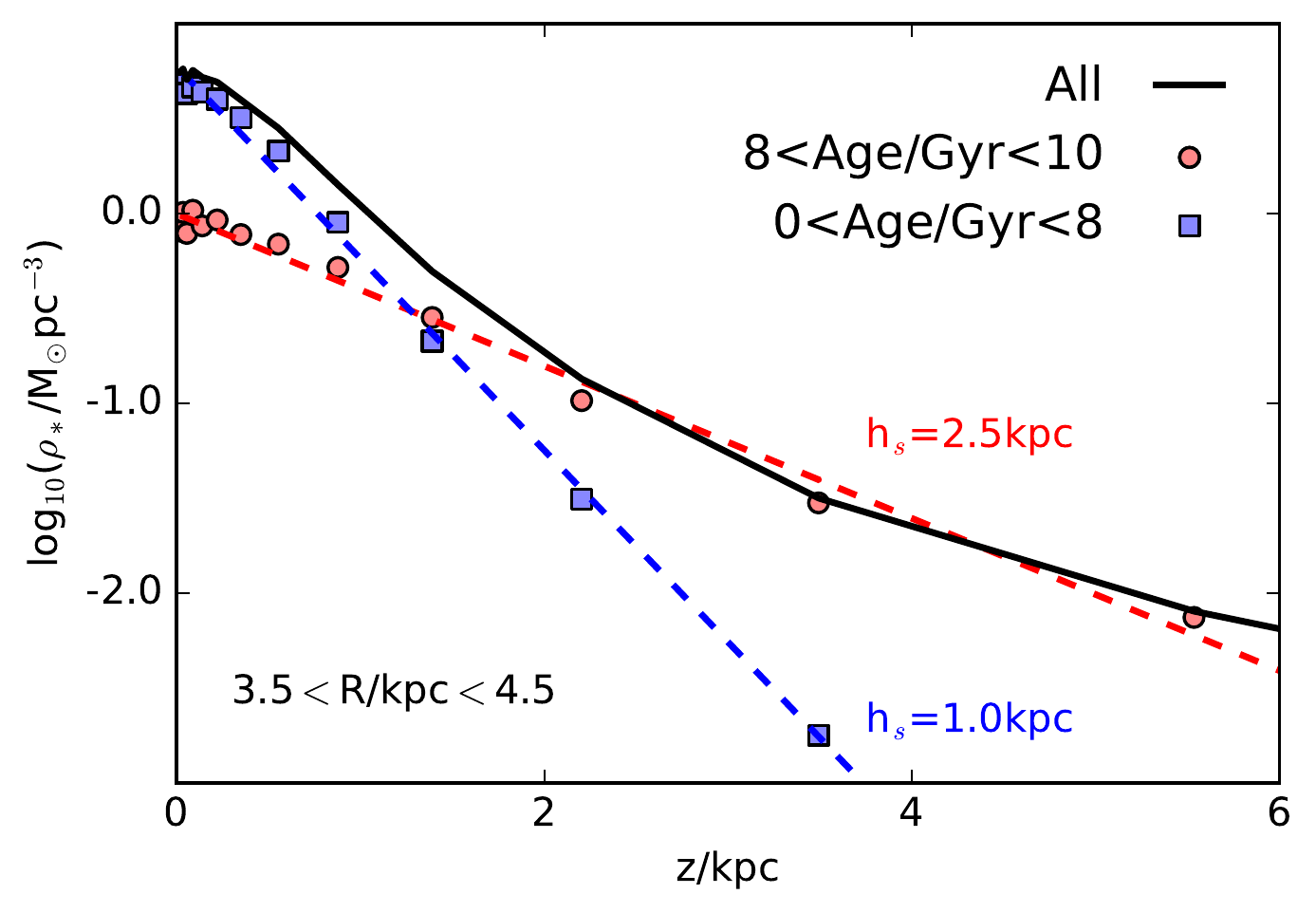}}\\%
  \caption{ Vertical density distribution, $\rho(z)$, for all disc
    stars with radii $3.5<R/$kpc$<4.5$ (black line). Old stars
    contribute mostly to a `thick' disc; young stars to a `thin'
    disc, each well approximated by a simple exponential law.}
  \label{FigThickDisc}
\end{figure}

We present in this brief appendix, for the sake of completeness and
reference, the circular velocity profile of the galaxy at various
times (Fig.~\ref{FigRVcirc}), as well as the vertical density profile
of star particles at a cylindrical radius $R=4$ kpc
(Fig.~\ref{FigThickDisc}). The former shows that our galaxy has a
reasonably flat circular velocity profile, as commonly observed for
spiral galaxies. The latter shows that, as for the Milky Way
\citep{Gilmore1983}, the vertical density profile of stars is not well
approximated by a simple exponential law, but can be described by the
superposition of a `thick' and a `thin' disc.

\bibliographystyle{mnras}
\bibliography{master}

\label{lastpage}
\end{document}